\documentclass[12pt]{article}

\usepackage{graphicx}

\newcommand{\PV}[1]{P\left({#1}\right)}

\newcommand{\eps}{\epsilon}

\newcommand{\pup}{p^{\uparrow}}

\newcommand{\et}{\widetilde{e}}

\newcommand{\xh}{\hat{x}}
\newcommand{\zh}{\hat{z}}
\newcommand{\sigmah}{\hat{\sigma}}
\newcommand{\xbj}{x_{bj}}

\newcommand{\eh}{\widehat{e}}
\newcommand{\Eh}{\widehat{E}}
\newcommand{\Et}{\widetilde{E}}

\newcommand{\Deltat}{\widetilde{\Delta}}
\newcommand{\St}{\widetilde{S}}

\newcommand{\lrarrow}{\leftrightarrow}

\newcommand{\nn}{\nonumber}

\newcommand{\bfc}{\begin{figure}\begin{center}}
\newcommand{\efc}{\end{center}\end{figure}}

\newcommand{\fig}[2]{\scalebox{#1}{\includegraphics{#2}}}

\newcommand{\Slash}[1]{\ooalign{\hfil/\hfil\crcr$#1$}}

\newcommand{\bra}[1]{\langle #1 |}
\newcommand{\ket}[1]{| #1 \rangle}

\newcommand{\Tr}[1]{{\rm Tr}\left[ #1 \right]}
\newcommand{\vb}{\biggl|}
\newcommand{\cl}{\biggr|_{\rm c.l.}}

\newcommand{\Oab}{\Omega^{\alpha}_{\,\,\,\,\beta}}
\newcommand{\psib}{\bar{\psi}}
\newcommand{\del}[2]{\frac{\partial #1}{\partial #2}}

\newcommand{\dZ}[1]{\frac{dz_{#1}}{z_{#1}^2}}
\newcommand{\lpartial}{\overleftarrow{\partial}}

\begin{document}

\vspace*{0.5cm}

\begin{center}

{\Large \bf Contribution of twist-3 fragmentation function \\ [3mm]
 to single transverse-spin asymmetry \\[5mm]
in semi-inclusive deep inelastic scattering}

\vspace{1cm}

Koichi Kanazawa$^{1,2}$ and Yuji Koike$^3$

\vspace{1cm}

{\it $^1$ Graduate School of Science and Technology, Niigata University,
Ikarashi, \\ Niigata 950-2181, Japan}

\vspace{0.5cm}

{\it $^2$ Department of Physics, Barton Hall, Temple University, Philadelphia, Pennsylvania 19122, USA}

\vspace{0.5cm}

{\it $^3$ Department of Physics, Niigata University,
Ikarashi, Niigata 950-2181, Japan}

\vspace{2.5cm}

{\large \bf Abstract} \end{center}

We study the contribution of the twist-3 fragmentation function to the
single-transverse spin asymmetry in semi-inclusive deep inelastic
scattering within the framework of the collinear factorization.
Making use of 
the Ward-Takahashi identity in QCD, we establish the 
formalism in the Feynman gauge to calculate the {\it non-pole}
contribution of the twist-3 fragmentation function to the asymmetry and derive
the complete cross-section
formula in the leading order QCD perturbation theory.  
The obtained twist-3
hadronic tensor is shown to satisfy the electromagnetic gauge-invariance.
The behavior in the small transverse-momentum region for the five structure functions
is also given.

\newpage

\section{Introduction}
Clarifying the origin of the single spin-asymmetries (SSA)
in hard inclusive processes has been a big challenge in high energy spin physics.
Since the asymmetries reflect the effect of parton's intrinsic transverse momentum and
correlations in the hadrons,
proper description of SSAs requires an extension of the theoretical frameworks
for high-energy processes beyond the conventional twist-2 level.   
Within the framework of the collinear factorization, which is
suited to describe the hadron production with a large transverse momentum $p_T$, 
SSAs appear as a twist-3 observable representing the multi-parton (quark-gluon or purely gluonic)
correlations in the nucleon and in the fragmentation
process.  
So far the formalism for calculating the contribution 
from the twist-3 distributions 
to SSA has been well 
developed\,\cite{ET1982,QS1992,EguchiKoikeTanaka2007,BeppuKoikeTanakaYoshida2010}, and
there have been many works on them.  
Those works for the quark-gluon correlation functions include SSAs
in semi-inclusive deep-inelastic scattering (SIDIS, 
$ep^\uparrow\to ehX$)\,\cite{EguchiKoikeTanaka2007,JiQiuVogelsangYuan2006SIDIS,
EguchiKoikeTanaka2006,KoikeTanaka2007M,KoikeVogelsangYuan2008,
KoikeTanaka2009}, 
hadron production in $pp$ collision 
($p^\uparrow p\to hX$)\,\cite{QiuSterman1998,KanazawaKoike2000,KanazawaKoike2000E,
KouvarisQiuVogelsangYuan2006,KoikeTanaka2007,KoikeTomita2009,KanazawaKoike2010,KanazawaKoike2011},
the Drell-Yan and direct-photon production ($p^\uparrow p\to \gamma^{(*)}X$)\,\cite{QS1992,JiQiuVogelsangYuan2006DY,
KanazawaKoike2011DY}, dijet production\,\cite{QiuVogelsangYuan2007}
and $W$-production\,\cite{Metz:2010xs} in $pp$ collision, 
and $ep\to {\rm jet} X$\,\cite{KangMetzQiuZhou2011}, etc.  
The works on the three-gluon correlation functions include those for
the $D$-meson production in SIDIS 
($ep^\uparrow\to eDX$)\,\cite{BeppuKoikeTanakaYoshida2010,KQ08,
KoikeTanakaYoshida2011,BeppuKoikeTanakaYoshida2012}
and $pp$-collision ($p^\uparrow p\to DX$)\,\cite{KQVY08,KoikeYoshida2011} and the Drell-Yan and direct-photon production
($p^\uparrow p\to \gamma^{(*)}X$)\,\cite{KoikeYoshida2012}, etc. 
It has been also shown that the 
partonic hard cross section for the soft-gluon-pole contribution from the twist-3 distributions
can be obtained from a certain twist-2 partonic hard cross section\,\cite{KoikeTanaka2007M,
KoikeTanaka2007,KoikeTanakaYoshida2011,KoikeYoshida2011,KoikeYoshida2012}.  This 
``master formula'' 
greatly simplifies the actual calculation, making the structure of the cross section transparent,
and is expected to be a useful tool to include higher order corrections. 
Despite these remarkable developments in our handling of the twist-3 distributions, 
study on the twist-3 fragmentation functions is still limited\,\cite{YuanZhou2009,KangYuanZhou2010,MetzPitonyak2013}.

In this paper, we shall derive the contribution
from the twist-3 fragmentation function to the single-spin-dependent cross section
for SIDIS, $ep^\uparrow\to ehX$.
In the case of the twist-3 distribution
functions, a pole part of an internal propagator in the partonic hard scattering amplitude
gives rise to the real cross sections, being combined with
the {\it real} twist-3 multi-parton distribution functions.  
On the other hand, the twist-3 fragmentation functions corresponding to the
quark-gluon correlation are, in general, complex because the time-reversal 
invariance does not give any constraint owing to the final state interaction\,\cite{Ji1994},
and thus the imaginary part of the fragmentation matrix elements
contributes to the cross section 
combined with a nonpole partonic hard-scattering part. 
One should also recall 
that the soft-gluon-pole matrix elements of the
twist-3 quark-gluon
fragmentation function has shown to be identically zero, which is connected
to the universality property (process-independence) of the Collins fragmentation function\,\cite{MeissnerMetz2009,
GambergMukherjeeMulders2011}.
Therefore we shall focus on the 
nonpole contribution of the twist-3 fragmentation in this work.
As in our previous study for
the twist-3 quark-gluon and three-gluon correlation functions\,\cite{EguchiKoikeTanaka2007,BeppuKoikeTanakaYoshida2010},
we will employ the Feynman-gauge to calculate the
cross section, since this formulation 
makes clear how the color-gauge-invariant fragmentation
matrix elements are constructed and factorized in the cross section.
The electromagnetic gauge invariance is also manifest in our formulation.

Here we mention previous works on the nonpole contributions
from the twist-3 fragmentation functions. 
Yuan and Zhou\,\cite{YuanZhou2009} focused on the small-$p_T$ behavior of the
twist-3 fragmentation contribution
to the Collins asymmetry in SIDIS in comparison with the 
transverse-momentum-dependent (TMD) factorization approach,
but did not present the twist-3 cross section.  
We will present the cross section formula for all five structure functions 
in $ep^\uparrow \to ehX$.  
We will also see that our leading small-$p_T$ behavior of the cross section for the Collins 
asymmetry disagrees with the result in \cite{YuanZhou2009}. 
Kang {\it et al.}\,\cite{KangYuanZhou2010} calculated the so-called ``derivative term"
of the contribution to $A_N$ of $p^\uparrow p\to hX$.  
Metz and Pitonyak\,\cite{MetzPitonyak2013} derived the complete leading-order formula for the
twist-3 fragmentation contribution to this process, using the 
lightcone-gauge formulation 
which was developed for the 
nonpole contribution from the twist-3 distribution
to other $p_T$-dependent processes\,\cite{Zhou:2009jm,Metz:2012fq,Liang:2012rb}. 
Our formulation in Feynman gauge as well as its application to
SIDIS will shed new light for the study on the twist-3 fragmentation functions. 
We also mention that 
one of the present authors (KK) applied the present Feynman-gauge formulation
for the nonpole contribution 
of the three-gluon correlation functions 
to $A_{LT}$ in $pp$-collision\,\cite{HattaKanazawaYoshida2013}.  
Together with the contribution from the twist-3 quark-gluon and the three-gluon correlation
functions derived in \cite{EguchiKoikeTanaka2007,KoikeTanaka2009,BeppuKoikeTanakaYoshida2010},
the present study will complete the twist-3 cross section for $ep^\uparrow\to ehX$
which is relevant for the future eRHIC experiment\,\cite{INT}.

The remainder of this paper is organized as follows:
In section 2, we define the twist-3 fragmentation functions for a spin-0 hadron
which are relevant in our analysis.  
Both 2-parton and 3-parton correlation functions contribute to $ep^\uparrow\to ehX$,
and all of them are chiral-odd, appearing in a pair with the transversity distribution from the polarized nucleon.  
In section 3, we will describe our Feynman-gauge formalism to
calculate the nonpole contribution from the twist-3 fragmentation function.  
To this end we apply the collinear expansion to the partonic hard scattering part.  
Use of the Ward-Takahashi (WT) identities allows us to reorganize all the contribution
into
a color-gauge-invariant factorized form.  
We also show the obtained hadronic tensor satisfies the electro-magnetic gauge invariance,
using the relation among twist-3 fragmentation functions.
In section 4, we will show the result for the five independent structure functions
corresponding to different azimuthal structures for $ep^\uparrow\to ehX$.  
In order to make connection with the TMD
factorization approach, we also show the small $p_T$-limit of 
each structure function, in comparison with those from the twist-3 distributions.  
Section 5 will be devoted to a brief summary.

\section{Twist-3 fragmentation functions}

In this section, we introduce twist-3 fragmentation functions
for a spin-0 hadron relevant to the present study.   
In the twist-3 accuracy, the four momentum of the hadron 
$P_h$ can be regarded as lightlike.  
The 2-quark correlator defines two {\it real} twist-3 fragmentation functions $ \widehat{e}_1 (z)$
and $\eh_{\bar{1}} (z)$ as 
\cite{Ji1994}
\begin{eqnarray}
\Delta_{ij} (z) &=& \frac{1}{N} \sum_X \int \frac{d\lambda}{2\pi}
 e^{-i\frac{\lambda}{z}} \bra{0} \psi_i(0) \ket{h(P_h)X} \bra{h(P_h)X}
 \psib_j(\lambda w) \ket{0} \nn\\
 &=& \frac{M_N}{z} ({\bf
1})_{ij} \widehat{e}_1 (z) + \frac{M_N}{2z}
  (\sigma_{\lambda\alpha}i\gamma_5)_{ij} \epsilon^{\lambda\alpha wP_h} 
  \eh_{\bar{1}} (z) + \cdots \label{2-body},
\end{eqnarray}
where $\psi_i$ is the quark field with spinor index $i$. Here and below
we use the nucleon mass
$M_N$ to define all the twist-3 fragmentation functions as dimensionless quantities.
\footnote{This convention is convenient, since 
all contributions from the twist-3 distribution (quark-gluon and three-gluon) and
fragmentation functions have a common factor in the cross section.} 
$w^\mu$ is a light-like vector satisfying $P_h\cdot w=1$.  
More specifically, for $P_h^\mu=(|\vec{P}_h|, \vec{P}_h)$,
$w^\mu=(|\vec{P}_h|, -\vec{P}_h)/(2|\vec{P}_h|^2)$. 
$N=3$ is the number of colors and $\eps^{\lambda\alpha w P_h} \equiv
\eps^{\lambda\alpha \rho\sigma} w_{\rho} P_{h\sigma}$ with the
Levi-Civita tensor being $\eps_{0123} \equiv +1$.  We have suppressed the
gauge-link operator between the fields for simplicity. 
$\eh_{\bar{1}} (z)$ is a naively $T$-odd function which can be nonzero due to
the final-state interaction and
could contribute to nonzero SSA, 
while $\hat{e}_{1}$ can not cause the asymmetry.   

We next introduce the $F$-type 
twist-3 fragmentation functions defined from the 
quark-gluon correlation functions\,\cite{EguchiKoikeTanaka2006}
\begin{eqnarray}
%
%
\Delta_{F ij}^{\alpha} (z_1,z_2) &=& \frac{1}{N} \sum_X
\int\frac{d\lambda}{2\pi} \frac{d\mu}{2\pi}
e^{-i\frac{\lambda}{z_1}} e^{-i\mu(\frac{1}{z_2}-\frac{1}{z_1})}
\bra{0} \psi_i (0) \ket{hX} \bra{hX} \bar{\psi}_j(\lambda
w)  gF^{\alpha\beta} (\mu w) w_{\beta} \ket{0} \nn\\
&=& \frac{M_N}{2z_2} (\gamma_5 \Slash{P}_h \gamma_{\lambda})_{ij}
\eps^{\lambda \alpha w P_h} \Eh_F (z_1,z_2) + \cdots, \label{Eh_F}\\
%
%
\tilde{\Delta}_{F ij}^{\alpha} (z_1,z_2) &=& \frac{1}{N} \sum_X \int\frac{d\lambda}{2\pi} \frac{d\mu}{2\pi}
e^{-i\frac{\lambda}{z_1}} e^{-i\mu(\frac{1}{z_2}-\frac{1}{z_1})}
\bra{0} \psib_j (\lambda w) \psi_i(0)
\ket{hX} \bra{hX} gF^{\alpha\beta} (\mu w) w_{\beta}
  \ket{0} \nn\\
&=& \frac{M_N}{2z_2} (\gamma_5 \Slash{P}_h \gamma_{\lambda})_{ij}
\eps^{\lambda \alpha w P_h} \Et_F (z_1,z_2) + \cdots, \label{Et_F}
\end{eqnarray}
where $F^{\alpha\beta}\equiv F^{a\alpha\beta}T^a$ is the gluon's field
strength with $T^a$ being the color matrices.
$\Eh_F$ and  $\Et_F$ are, in general, complex functions.  
$\Eh_F(z_1,z_2)$ has a support on $1>z_2>0$ and $z_1>z_2$, while 
$\Et_F(z_1,z_2)$ has a support on ${1\over z_2}-{1\over z_1}>1$, ${1\over z_1}<0$
and ${1\over z_2}>0$.  
%
Replacing the field strength tensor $gF^{\alpha\beta} (\mu w) w_\beta$
by the covariant derivative $D^\alpha (\mu w)=\partial^\alpha -igA^\alpha (\mu w)$ 
in (\ref{Eh_F}), one can define (complex) $D$-type function $\Eh_D (z_1,z_2)$ by the same 
decomposition in the right-hand-side. However, $\Eh_D (z_1,z_2)$ is related to 
$\Eh_F(z_1,z_2)$ as\,\cite{EguchiKoikeTanaka2006}
\begin{eqnarray}
\Eh_D (z_1,z_2) = P \left( \frac{1}{1/z_1-1/z_2} \right)
\Eh_F (z_1,z_2) + 
\delta\left(\frac{1}{z_1}-\frac{1}{z_2} \right) \et (z_2) \label{relation}, 
\end{eqnarray}
where $\et (z)$ is given by
%
\begin{eqnarray}
 \Delta_{\partial ij}^\alpha (z) &=& \frac{1}{N} \sum_X \int
  \frac{d\lambda}{2\pi} e^{ -i\frac{\lambda}{z} } \bra{0} [\infty w,
  0] \psi_i(0)
   \ket{hX} \bra{hX} \bar{\psi}_j (\lambda w)
  [\lambda w,\infty w]
  \ket{0} \lpartial^\alpha \nn\\
 &=& \frac{M_N}{2z} (\gamma_5 \Slash{P_h}\gamma_\lambda)_{ij}
  \epsilon^{\lambda\alpha wP_h} \et (z) + \cdots. \label{Deltapartial}
\end{eqnarray}
Here we have
restored the gauge-link operators to emphasize that the derivative
$\lpartial^\alpha$ hits both $\psib(\lambda w)$ and $[\lambda w, \infty w]$,
and in writing down the relation (\ref{relation})
we have used the identity 
$\Eh_F(z,z)=0~$\cite{MeissnerMetz2009,GambergMukherjeeMulders2011}.
The function $\et (z)$ can be shown to be purely imaginary, i.e., ${\rm Im}\, \et (z)= -i\,  \et (z)$. 
From 
QCD equation of motion, one has the relation \cite{MetzPitonyak2013},
\begin{eqnarray}
 z\int \frac{dz'}{z'^2} \Eh_D(z',z) &=& \eh_1(z) +
  i\eh_{\bar{1}}(z), \label{EOM}
\end{eqnarray}
which, combined with the relation (\ref{relation}), 
leads to an important relation in our analysis
\begin{eqnarray}
z\int {dz'\over z'^2} P \left( \frac{1}{1/z'-1/z} \right)
{\rm Im}\,\Eh_F (z_1,z_2) + z{\rm Im}\, \et (z) 
=\eh_{\bar{1}}(z).  
\label{EOM2}
\end{eqnarray}
In what follows, we shall obtain the
contribution of the twist-3 fragmentation function
to the single-spin-dependent cross section
in terms of ${\rm Im}\,\Eh_F (z_1,z_2)$, ${\rm Im}\, \et (z)$ and
$\eh_{\bar{1}}(z)$.  But the relation (\ref{EOM2}) allows us to
eliminate $\eh_{\bar{1}}(z)$ (or ${\rm Im}\, \et (z)$) in favor of the other two functions.

\section{Collinear twist-3 formalism in Feynman gauge}

\subsection{Kinematics}

Following\,\cite{KoikeTanakaYoshida2011,MengOlnessSoper1992}, we summarize the
kinematics for the SIDIS process,
\begin{eqnarray}
e (l) + \pup (p, S) \to e(l') + h(P_h) + X,
\label{SIDIS}
\end{eqnarray}
where $l$, $l'$ and $P_h$ are 4-momenta of the initial lepton, scattered
lepton, and the produced hadron, respectively.
This process is described by the five independent Lorentz invariants: 
\begin{eqnarray}
&& S_{ep} = (p+l)^2 ,\quad \xbj = \frac{Q^2}{2p\cdot q} ,\quad Q^2 =
  -q^2 = -(l-l')^2, \nn\\
&& z_f = \frac{p\cdot P_h}{p\cdot q} ,\quad q_T = \sqrt{-q_t^2},  
\end{eqnarray}
where the space-like momentum $q_t$ is the ``transverse'' component of $q$ 
defined as
\begin{eqnarray}
q_t^\mu= q^\mu - {P_h\cdot q\over p\cdot P_h} p^\mu - {p\cdot q\over
 p\cdot P_h} P_h^\mu,
\end{eqnarray}
satisfying 
$q_t\cdot p=q_t\cdot P_h=0$.  
In the hadron frame
where the virtual photon and the initial nucleon are
collinear, i.e., both move along the $z$-axis,
the momenta $q$ and $p$ are given by
\begin{eqnarray}
 q^\mu &=& (q^0, \vec{q}) = (0,0,0,-Q), \nn\\
 p^\mu &=& \left( \frac{Q}{2\xbj}, 0, 0, \frac{Q}{2\xbj} \right). \nn
\end{eqnarray}
The azimuthal angle of the hadron plane as measured from the $xz$ plane
is taken to be
$\chi$ and thus the momentum of the final hadron is parameterized as
\begin{eqnarray}
P_h^\mu = {z_f Q \over 2}\left( 1 + {q_T^2\over Q^2}, {2 q_T\over Q}\cos\chi,
{2 q_T\over Q}\sin\chi,
-1+{q_T^2\over Q^2} \right) \; .
\label{Dmomem}
\end{eqnarray}
Correspondingly $w^\mu$ becomes
\begin{eqnarray}
w^\mu = { 1 \over z_f Q \left( 1 +{q_T^2\over Q^2}\right)^2 }
\left( 1 + {q_T^2\over Q^2}, -{2 q_T\over Q}\cos\chi,
-{2 q_T\over Q}\sin\chi,
1-{q_T^2\over Q^2} \right) \; .
\label{wmu}
\end{eqnarray}
The transverse momentum of the final hadron in this 
frame is given by $P_{hT}=z_f q_T$, which is true in any frame where the
3-momenta $\vec{q}$ and $\vec{p}$ 
are collinear.  The azimuthal angle of the lepton plane measured from the $xz$ plane is taken to be $\phi$ and 
thus the lepton momentum can be parameterized as
\begin{eqnarray}
l^\mu & = & \frac{Q}{2}\left( \cosh\psi,\sinh\psi\cos\phi,
\sinh\psi\sin\phi,-1\right),\nn\\
l'^\mu & = & \frac{Q}{2}\left( \cosh\psi,\sinh\psi\cos\phi,
\sinh\psi\sin\phi,1\right),
\label{eq2.lepton}
\end{eqnarray}
with $\cosh\psi = {2\xbj S_{ep}\over Q^2} -1$.
The transverse spin vector of the initial nucleon $S^\mu$ is
parametrized as
\begin{eqnarray}
S^\mu= (0,\cos\Phi_S,\sin\Phi_S,0),
\label{phis}
\end{eqnarray}
with the azimuthal angle $\Phi_S$ of $\vec{S}$.  
Although three azimuthal angles $\phi$, $\chi$ and $\Phi_S$ are defined above, 
it is obvious that 
the cross section for $ep^\uparrow\to eh X$ depends on them only through the relative angles 
$\phi-\chi$ and $\Phi_S-\chi$.  
Thus, it can be expressed in terms of
$S_{ep}$, $\xbj$, $Q^2$, $z_f$, $q_T^2$, $\phi-\chi$ and $\Phi_S-\chi$ in the above hadron
frame.
Note that $\phi$, $\chi$ and $\Phi_S$ are invariant under boosts in the 
$\vec{q}$-direction, so that the cross section presented below is
the same in any frame where $\vec{q}$ and $\vec{p}$ are collinear.  

With the kinematical variables defined above, 
the differential cross section for 
the SIDIS process with the unpolarized lepton
is expressed as 
\begin{eqnarray}
\frac{d^6 \Delta \sigma}{d \xbj dQ^2 dz_f dq_T^2 d\phi d\chi}
 = \frac{\alpha_{em}^2}{128\pi^4 \xbj^2 S_{ep}^2 Q^2}
 z_f L^{\mu\nu}(l, l')W_{\mu\nu}(p,q,P_h),
\label{diffsigma}
\end{eqnarray}
where $L^{\mu\nu}(l, l')=2(l^\mu l'^\nu + l^\nu l'^\mu)-Q^2g^{\mu\nu}$
is the leptonic tensor,
$W_{\mu\nu}(p,q,P_h)$ is the hadronic tensor in the same normalization as in \cite{BeppuKoikeTanakaYoshida2010}, and  
$\alpha_{em}=e^2/(4\pi)$ is the QED coupling constant.  
The single-spin-dependent part in the cross section (\ref{diffsigma})
describes the SSA in the process (\ref{SIDIS}).

\subsection{Analysis of hadronic tensor}

\begin{figure}[t]
 \begin{center}
  \fig{0.5}{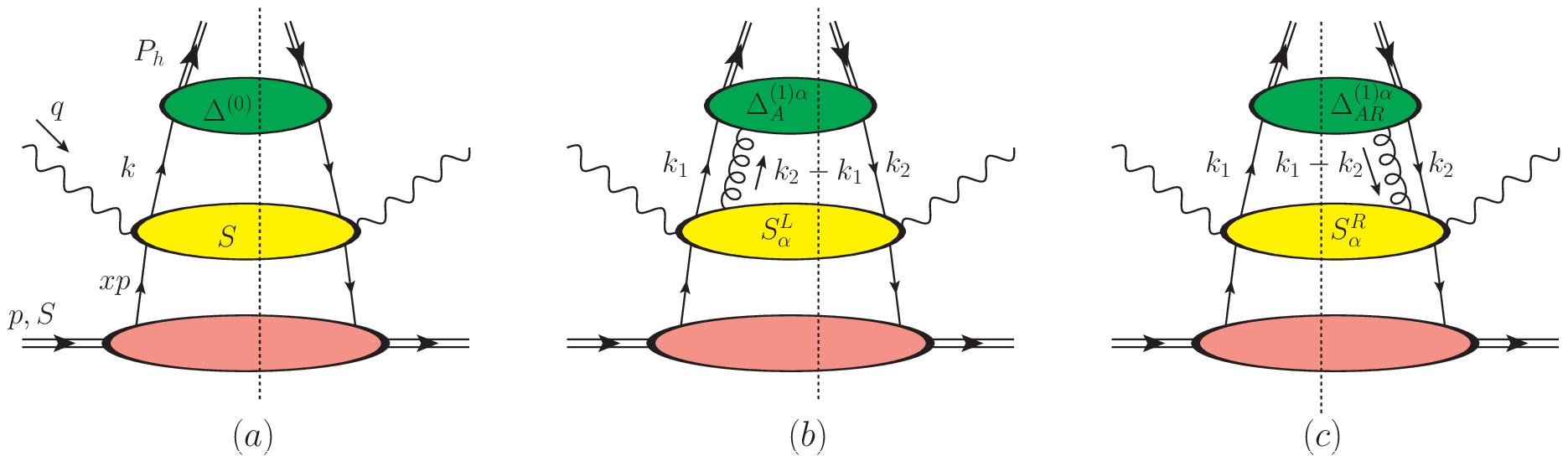}
  \fig{0.5}{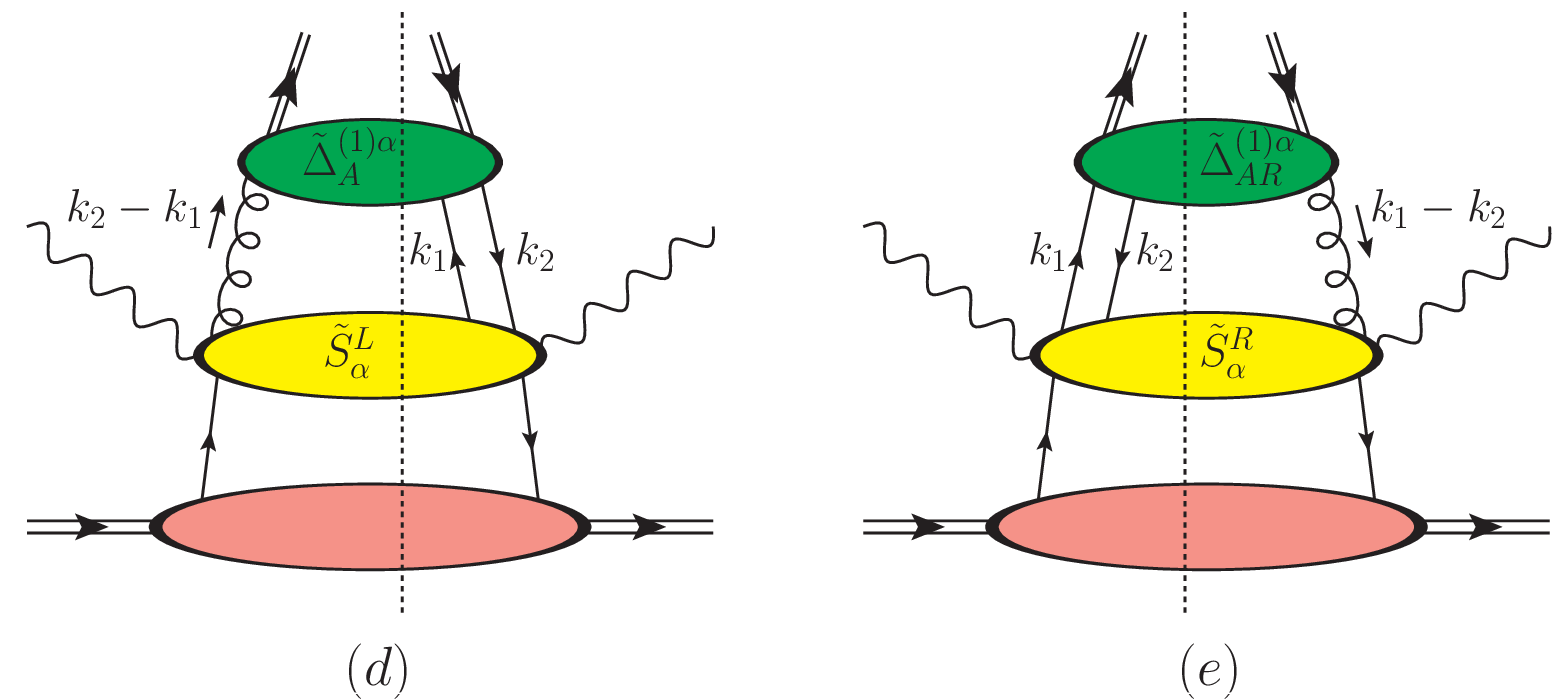}
  \caption{Generic diagrams giving rise to the twist-3 fragmentation
  function contribution to the polarized cross section $\Delta\sigma$.
The upper blob represents the fragmentation matrix elements for the final hadron
and the lower blob is the quark-transversity distribution.  The middle blob
denotes the partonic hard scattering part.}
\label{generic}
 \end{center}
\end{figure}

We now consider the contribution from the twist-3 fragmentation function to the hadronic tensor $W_{\mu\nu} (p, q, P_h)$.  
Figure \ref{generic} shows the generic diagrams which give rise to
the twist-3 effect originating from the final-state hadron.
In these contributions, the quark transversity distribution $h(x)$ can be 
immediately factorized from the hadronic
tensor as
\begin{eqnarray}
W_{\mu\nu} (p, q, P_h) = \int \frac{dx}{x} h (x) w_{\mu\nu} (xp,
q, P_h),
\end{eqnarray}
where $x$ denotes the momentum fraction of the quark
in the nucleon, and the summation over quark flavors is implicit.
$w_{\mu\nu}$ receives contribution from each of
the five diagrams depicted in Fig.~\ref{generic} as
\begin{eqnarray}
w_{\mu\nu} &\equiv& w^{(a)}_{\mu\nu} + w^{(b)}_{\mu\nu} + w^{(c)}_{\mu\nu} + w^{(d)}_{\mu\nu} + w^{(e)}_{\mu\nu} \nn\\
&\equiv& \int \frac{d^4k}{(2\pi)^4} \Tr{
\Delta^{(0)} (k) S_{\mu\nu} (k)  } \nn\\
&& + \int \frac{d^4k_1}{(2\pi)^4} \frac{d^4k_2}{(2\pi)^4} 
\left\{ 
\Tr{ \Delta^{(1)\alpha}_A (k_1,k_2) S^L_{\mu\nu,\alpha} (k_1,k_2) } 
+ \Tr{ \Delta^{(1)\alpha}_{AR} (k_1,k_2) S^R_{\mu\nu,\alpha} (k_1,k_2) } 
\right\} \nn\\
&&  + \int \frac{d^4k_1}{(2\pi)^4} \frac{d^4k_2}{(2\pi)^4} 
\left\{	\Tr{ \widetilde{\Delta}^{(1)\alpha}_A (k_1,k_2) \widetilde{S}^L_{\mu\nu,\alpha}
 (k_1,k_2) } + 
+ \Tr{ \widetilde{\Delta}^{(1)\alpha}_{AR} (k_1,k_2) \widetilde{S}^R_{\mu\nu,\alpha}
 (k_1,k_2) } \right\},\nn\\
\label{DE}
\end{eqnarray}
where $k$ and $k_{1,2}$ are the 4-momenta of the quarks fragmenting into
the final hadron, and we have suppressed the other momentum arguments
$P_h$, $q$ and $p$ for simplicity.  
$S_{\mu\nu}(k)$, $S^{L,R}_{\mu\nu,\alpha} (k_1,k_2)$ and $\St^{L,R}_{\mu\nu,\alpha} (k_1,k_2)$ are the
partonic hard parts represented by the
middle blobs in Fig.\,\ref{generic}.   
Tr$[\cdots]$ indicates the trace over spinor indices while the color trace is implicit.  
The hadronic matrix elements $\Delta^{(0)}(k)$, $\Delta^{(1)\alpha}_{A}(k_1,k_2)$
and $ \Deltat^{(1)\alpha}_{A}(k_1,k_2)$ in
Eq.~(\ref{DE}) corresponding to Fig.\,\ref{generic} (a), (b) and (d), respectively, are defined as
\begin{eqnarray}
\hspace{-0.7cm}
\Delta^{(0)}_{ij} (k) &=& \frac{1}{N} \sum_X \int d^4\xi e^{-ik\cdot \xi}
\bra{0} \psi_i (0) \ket{h X} \bra{h X} \bar{\psi}_j (\xi) \ket{0}, 
\label{Delta}\\
\hspace{-0.7cm} \Delta^{(1)\alpha}_{Aij} (k_1,k_2) &=& \frac{1}{N} \sum_X \int d^4\xi
\int d^4\eta e^{-ik_1\cdot \xi} e^{-i(k_2-k_1)\cdot\eta} 
\bra{0} \psi_i (0) \ket{h X} \bra{h X} \bar{\psi}_j (\xi)
 gA^{\alpha}
(\eta) \ket{0}, \label{DeltaA}
\\
\hspace{-0.7cm} \Deltat^{(1)\alpha}_{Aij} (k_1,k_2) &=&  \frac{1}{N} \sum_X \int d^4\xi
 \int d^4\eta e^{-ik_1\cdot \xi} e^{-i(k_2-k_1)\cdot\eta} 
\bra{0} \psi_i (0) \bar{\psi}_j (\xi) \ket{h X} \bra{h X} gA^{\alpha}
(\eta) \ket{0}.\label{Deltatilde}
\end{eqnarray}
Here the numbers in the superscripts denote the number of the coherent gluon lines 
coming out of the fragmentation matrix elements.  
$\Delta^{(1)\alpha}_{AR}(k_1,k_2)$ for the diagram Fig.\,\ref{generic}(c)
is obtained from $\Delta^{(1)\alpha}_{A}(k_1,k_2)$ 
by shifting the gluon field $gA^{\alpha}(\eta)$ in 
(\ref{DeltaA}) to the right of the cut.  
Likewise,  $\Deltat^{(1)\alpha}_{AR} (k_1,k_2)$ for the diagram Fig.\,\ref{generic}(e)
is obtained from $ \Deltat^{(1)\alpha}_{A} (k_1,k_2)$
by the interchange of $\psi_i (0) \bar{\psi}_j (\xi)$ and $gA^{\alpha}(\eta)$
in (\ref{Deltatilde}).  
We note the relations among these matrix elements and the partonic hard parts,
$\Delta^{(1)\alpha}_{AR}(k_1,k_2)= \gamma^0 \Delta^{(1)\alpha}_{A} (k_2,k_1)^\dagger \gamma^0$,  
$\Deltat^{(1)\alpha}_{AR}(k_1,k_2)= 
\gamma^0 \Deltat^{(1)\alpha}_{A} (k_2,k_1)^\dagger \gamma^0$,  
$S^R_{\mu\nu,\alpha} (k_1,k_2)=\gamma^0 S^L_{\nu\mu,\alpha} (k_2,k_1)^\dagger \gamma^0$ and
$\widetilde{S}^R_{\mu\nu,\alpha} (k_1,k_2)=\gamma^0 \widetilde{S}^L_{\nu\mu,\alpha} (k_2,k_1)^\dagger \gamma^0$.  
These relations guarantees that the cross section is real.  
Since the leptonic tensor is symmetric (and real) under $\mu\leftrightarrow \nu$ 
for the unpolarized electron,
only the symmetric parts of $S_{\mu\nu}$ and $S_{\mu\nu,\alpha}^{L,R}$ contribute to the cross section.  
Therefore, in what follows, 
we shall omit the Lorentz indices $\mu$ and $\nu$ from $S_{\mu\nu}(k)$ and $S_{\mu\nu,\alpha}^{L,R}(k_1,k_2)$
and represent their symmetric parts as $S(k)$ and $S_{\alpha}^{L,R}(k_1,k_2)$, respectively, for simplicity.
Correspondingly, we simply write $w^{(a,b,c,d,e)}$ for $w_{\mu\nu}^{(a,b,c,d,e)}$.

In order to extract twist-3 effects 
we perform the collinear expansion of the hard part with respect to $k$ and $k_{1,2}$
around the momentum $P_h$. 
In the present study it is useful to define a projection tensor
\begin{eqnarray}
\Oab = g^{\alpha}\,_{\beta} - P_h^{\alpha} w_{\beta}.
\end{eqnarray}
With this tensor the momentum of a quark is expressed as
$k^{\alpha}=(k\cdot w) P_h^{\alpha} + \Oab k^{\beta}$.
Accordingly the collinear expansion of $S(k)$ may be performed as
\begin{eqnarray}
S (k) = S (z) + \del{S
(k)}{k^{\alpha}} \cl \Oab k^{\beta} + \cdots
\end{eqnarray}
where $k\cdot w = 1/z$ and we have used the short-hand notation
as $S(z)\equiv S(k=\frac{P_h}{z})$.  Here and below ``c.l.'' indicates the collinear 
limit $k\to P_h/z$. 
Having this expansion, the first term in ($\ref{DE}$) becomes up to
twist-3
\begin{eqnarray}
w^{(a)} &=& \int \frac{dz}{z^2} \Tr{\Delta^{(0)} (z) S
(z)} 
 - i \Oab \int \frac{dz}{z^2}
 \Tr{ \Delta_{\partial}^{(0)\beta} (z)
 \del{S(k)}{k^{\alpha}} \cl } \label{S},
\end{eqnarray}
with the light-cone matrix elements
\begin{eqnarray}
&& \Delta^{(0)}_{ij} (z) = \frac{1}{N} \sum_X \int \frac{d\lambda}{2\pi}
e^{-i\frac{\lambda}{z}} \bra{0} \psi_i (0) \ket{hX} \bra{hX} \bar{\psi}_j
(\lambda w) \ket{0}, \\
&& \Delta^{(0) \alpha}_{\partial ij} (z) = \frac{1}{N} \sum_X \int
\frac{d\lambda}{2\pi}
e^{-i\frac{\lambda}{z}} \bra{0} \psi_i (0) \ket{hX}
\bra{hX} \bar{\psi}_j (\lambda w) \lpartial^{\alpha} \ket{0} \label{D2}.
\end{eqnarray}
where we have used a short-hand notation
\begin{eqnarray}
\partial^\alpha f(\lambda w) = \del{f(\xi)}{\xi_\alpha}
\vb_{\xi = \lambda w}.
\end{eqnarray}

Next, we consider the twist-3 effects arising from the quark-gluon
correlation matrix elements in $w^{(b,c)}$.
As was the case for the contribution from the twist-3 distribution\,\cite{EguchiKoikeTanaka2007}, we need to
reorganize the twist-3 terms appearing in the collinear expansion to construct
gauge-invariant $F$-type matrix elements.
A main difference for the present case is that the nonperturbative matrix elements differ
between $w^{(b)}$ and $w^{(c)}$.  
Since the coupling of the transversely polarized gluon
$A_\perp^\sigma$\,\footnote{Here $\perp$ indicates the transverse component
with respect to $P_h$.} onto $S^{L}_{\sigma}(k_1,k_2)$
gives rise to a power suppressed contribution compared with the coupling of the 
longitudinally polarized gluon $(A\cdot w)P_h^{\sigma}$,
one needs collinear expansion only for the latter. 
For convenience we introduce the notation 
\begin{eqnarray}
S^L (k_1,k_2) \equiv P_h^{\sigma} S^L_{\sigma} (k_1,k_2)
\end{eqnarray}
and perform the collinear expansion for this quantity as
\begin{eqnarray}
S^L (k_1,k_2) &=& S^L (z_1,z_2)
+ \del{S^L (k_1,k_2)}{k_1^{\alpha}} \cl
\Oab k_1^{\beta} \nn\\
&& + \del{S^L (k_1,k_2)}{k_2^{\alpha}}
\cl \Oab k_2^{\beta} + \cdots. 
\end{eqnarray}
Substituting this expression into $w^{(b)}$, one finds in the twist-3 accuracy
\begin{eqnarray}
w^{(b)} &=& \int \frac{dz_1}{z_1^2} \frac{dz_2}{z_2^2}
\Tr{\Delta_A^{(1)\sigma} (z_1,z_2) w_{\sigma} S^L(z_1,z_2)} \nn\\
&& + \Oab \int \frac{dz_1}{z_1^2} \frac{dz_2}{z_2^2}
\Tr{ \Delta_A^{(1)\beta} (z_1,z_2) S^L_{\alpha} (z_1,z_2)
} \nn\\
&& - i\Oab \int \frac{dz_1}{z_1^2} \frac{dz_2}{z_2^2}
\Tr{ \Delta^{(1)\beta}_{\partial 1} (z_1,z_2)
\del{S^L(k_1,k_2)}{k_1^{\alpha}}
\cl } \nn\\
&& - i\Oab \int \frac{dz_1}{z_1^2} \frac{dz_2}{z_2^2}
\Tr{ \Delta^{(1)\beta}_{\partial 2} (z_1,z_2)
\del{S^L(k_1,k_2)}{k_2^{\alpha}}
\cl }, \label{wb2}
\end{eqnarray}
where $S^L_\sigma (z_1,z_2)\equiv S^L_\sigma (P_h/z_1,P_h/z_2)$, 
${\rm c.l.}$ indicates the collinear limit $k_i\to P_h/z_i$ ($i=1,\,2$), 
and the light-cone matrix elements are given by
\begin{eqnarray}
\Delta^{(1)\beta}_{Aij}(z_1,z_2) &=& \frac{1}{N} \sum_X \int \frac{d\lambda}{2\pi}
\frac{d\mu}{2\pi} e^{-i\frac{\lambda}{z_1}}
e^{-i(\frac{1}{z_2}-\frac{1}{z_1}) \mu}
\bra{0} \psi_i(0) \ket{hX} \nn\\
&& \times \bra{hX} \bar{\psi}_j (\lambda w) gA^{\beta} (\mu
w) \ket{0} ,\\
\Delta^{(1)\beta}_{\partial 1 ij}(z_1,z_2) &=& \frac{1}{N} \sum_X \int \frac{d\lambda}{2\pi}
\frac{d\mu}{2\pi} e^{-i\frac{\lambda}{z_1}}
e^{-i(\frac{1}{z_2}-\frac{1}{z_1}) \mu}
\bra{0} \psi_i(0) \ket{hX} \nn\\
&& \times \bra{hX} 
\bar{\psi}_j (\lambda w) \lpartial^{\beta} gA^w (\mu w) \ket{0} ,
\label{Deltapartial(1)}\\
\Delta^{(1)\beta}_{\partial 2 ij}(z_1,z_2) &=& \frac{1}{N} \sum_X \int
\frac{d\lambda}{2\pi}
\frac{d\mu}{2\pi} e^{-i\frac{\lambda}{z_1}}
e^{-i(\frac{1}{z_2}-\frac{1}{z_1}) \mu}
\bra{0} \psi_i(0) \ket{hX} \nn\\
&& \times  \bra{hX}  \bar{\psi}_j
(\lambda w) \lpartial^{\beta} gA^w (\mu w) + \bar{\psi}_j
(\lambda w) g (\partial^{\beta} A^w (\mu w)) \ket{0}.  
\end{eqnarray}
Note the first term in (\ref{wb2}) may be rewritten by the use of the
tree-level Ward identity
\begin{eqnarray}
P_h^{\sigma} S^L_{\sigma} (z_1,z_2) = \frac{S
(z_2)}{1/z_2-1/z_1+i\eps}. \label{GL}
\end{eqnarray}
By performing the integration over $1/z_1$, 
it is straightforward to see this term eventually becomes
$O(g)$ contribution of the gauge-link operator $[\lambda
w,\infty w]$ for $\Delta(z)$ in (\ref{2-body}).  The remaining terms in (\ref{wb2}) can
be reorganized as  
\begin{eqnarray}
w^{(b)} &=& 
 - i\Oab \int \frac{dz_1}{z_1^2} \frac{dz_2}{z_2^2} 
\Tr{ \Delta^{(1)\beta}_{\partial 1} (z_1,z_2) 
\left( \del{S^L(k_1,k_2)}{k_1^{\alpha}} 
\cl + \del{S^L(k_1,k_2)}{k_2^{\alpha}} 
\cl \right) } \nn\\
&& - i\Oab \int \frac{dz_1}{z_1^2} \frac{dz_2}{z_2^2}
\Tr{ (\Delta^{(1)\beta}_{\partial 2}(z_1,z_2) -
\Delta^{(1)\beta}_{\partial 1}(z_1,z_2)) 
\del{S^L(k_1,k_2)}{k_2^{\alpha}}
\cl } \nn\\
&& + \Oab \int \frac{dz_1}{z_1^2} \frac{dz_2}{z_2^2}
\Tr{ \Delta_A^{(1)\beta} (z_1,z_2) S^L_{\alpha} (z_1,z_2) }.
\end{eqnarray}
By writing $\Delta_{\partial 2}^{(1)} - \Delta_{\partial 1}^{(1)} =
\Delta_F^{(1)} + \widehat{\Delta}^{(1)}_A$ with 
\begin{eqnarray}
\Delta^{(1)\beta}_{Fij} &=& \frac{1}{N} \sum_X \int
\frac{d\lambda}{2\pi} \frac{d\mu}{2\pi} e^{-i\frac{\lambda}{z_1}}
e^{-i(\frac{1}{z_2} - \frac{1}{z_1}) \mu} \bra{0} \bar{\psi}_i (0) \ket{hX}
\nn\\
& &\qquad\qquad \times\bra{hX} \bar{\psi}_j(\lambda w) g \left(
\partial^{\beta} A^{w} (\mu w) -
w\cdot \partial A^{\beta} (\mu w) \right)
\ket{0},\label{DeltaF(1)}\\
\widehat{\Delta}^{(1)\beta}_{Aij} &=& \frac{1}{N} \sum_X \int
\frac{d\lambda}{2\pi} \frac{d\mu}{2\pi} e^{-i\frac{\lambda}{z_1}}
e^{-i(\frac{1}{z_2} - \frac{1}{z_1}) \mu} \bra{0} \bar{\psi}_i (0) \ket{hX}
\bra{hX} \bar{\psi}_j(\lambda w) g \partial^{\sigma} A^{\beta} (\mu w)
w_{\sigma} \ket{0}
\nn\\
&=& i\left( \frac{1}{z_2} - \frac{1}{z_1} \right) \Delta_{Aij}^{(1)\beta},
\end{eqnarray}
we find for the diagram (b) in Fig.\,\ref{generic}
\begin{eqnarray}
 w^{(b)} &=& -i\Oab \int \frac{dz_1}{z_1^2} \frac{dz_2}{z_2^2}
\Tr{ \Delta_F^{(1)\beta} (z_1,z_2) \del{S^L (k_1,k_2)}{k_2^{\alpha}}
\cl } \nn\\
&& + \Oab \int \frac{dz_1}{z_1^2} \frac{dz_2}{z_2^2}
\Tr{ \Delta_A^{(1)\beta} (z_1,z_2) \left(
\left(\frac{1}{z_2}-\frac{1}{z_1}\right)
\del{S^L (k_1,k_2)}{k_2^{\alpha}} \cl +
S^L_{\alpha} (z_1,z_2) \right) } \nn\\
&& -i \Oab \int \frac{dz_1}{z_1^2} \frac{dz_2}{z_2^2}
\Tr{ \Delta_{\partial 1}^{(1)\beta} (z_1,z_2) \left(
\del{S^L (k_1,k_2)}{k_1^{\alpha}} \cl +
\del{S^L (k_1,k_2)}{k_2^{\alpha}} \cl \right) }. \label{wb}
\end{eqnarray}
Carrying out a similar analysis for the diagram (c)
in Fig.\,\ref{generic}, one 
obtains
\begin{eqnarray}
 w^{(c)} &=& -i\Oab \int \frac{dz_1}{z_1^2} \frac{dz_2}{z_2^2}
\Tr{ \Delta_{FR}^{(1)\beta} (z_1,z_2) \del{S^R (k_1,k_2)}{k_2^{\alpha}}
\cl } \nn\\
&& + \Oab \int \frac{dz_1}{z_1^2} \frac{dz_2}{z_2^2}
\Tr{ \Delta_{AR}^{(1)\beta} (z_1,z_2) \left(
\left(\frac{1}{z_2}-\frac{1}{z_1}\right)
\del{S^R(k_1,k_2)}{k_2^{\alpha}} \cl +
S^R_{\alpha} (z_1,z_2) \right) } \nn\\
&& -i \Oab \int \frac{dz_1}{z_1^2} \frac{dz_2}{z_2^2}
\Tr{ \Delta_{\partial 1R}^{(1)\beta} (z_1,z_2) \left(
\del{S^R (k_1,k_2)}{k_1^{\alpha}} \cl +
\del{S^R (k_1,k_2)}{k_2^{\alpha}} \cl \right) } ,\label{wc}
\end{eqnarray}
where $\Delta_{\partial 1R}^{(1)\beta}$ and 
$\Delta_{FR}^{(1)\beta}$ are the correlation functions obtained by shifting the gluon fields
from the left of the cut to the right of the cut, respectively, 
in (\ref{Deltapartial(1)}) and (\ref{DeltaF(1)}), and are connected to
$\Delta_{\partial 1}^{(1)\beta}$ and 
$\Delta_{F}^{(1)\beta}$ by the relation 
$\Delta_{\partial 1R}^{(1)\beta}(z_1,z_2)=
 \gamma^0 \Delta_{\partial 1}^{(1)\beta}(z_2,z_1)^\dagger 
\gamma^0$ and 
$\Delta_{FR}^{(1)\beta}(z_1,z_2)=
 \gamma^0 \Delta_{F}^{(1)\beta}(z_2,z_1)^\dagger 
\gamma^0$.  
In our lowest order calculation with respect to the number of the coherent gluon lines, $\Delta_{F}^{(1)\beta}$
can be regarded as the $F$-type correlation
function defined in (\ref{Eh_F}).  
Therefore, 
except for the first terms in $w^{(a,b,c)}$, 
each term in
$w^{(a,b,c)}$ is not
gauge invariant.  
For the case of the pole contributions for the twist-3 distribution functions,
sum of the diagrams corresponding to the same type of poles
satisfies a particular WT identity due to the on-shell condition
for an internal line.  
This leads to a cancellation among all gauge-noninvariant terms,
leaving the gauge-invariant factorized expression for the 
corresponding twist-3 cross section\,\cite{EguchiKoikeTanaka2007,BeppuKoikeTanakaYoshida2010}.
In the present case, however, such special on-shell condition is
lacking, so that
one has to reconsider the factorization
property and prove the color gauge-invariance of the contribution of the
twist-3 fragmentation functions.
In the next subsection, we present such argument
 to establish the collinear twist-3 formalism for the non-pole
 contribution.

\subsection{Ward-Takahashi identities and relation among hard parts}
\begin{figure}
 \begin{center}
  \fig{0.7}{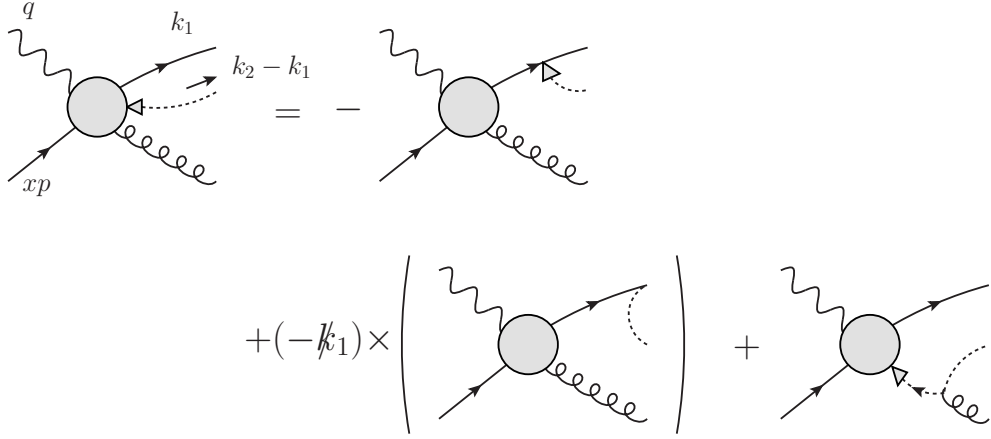}
 \end{center}
 \caption{WT identity for a coupling of the
 scalar-polarized gluon with the momentum $(k_1-k_2)^\sigma$ onto the amplitude in the
 left of the
 final-state cut of $S^{La}_\sigma (k_1,k_2)$. 
For the quark line with the on-shell momentum $xp$,
the spinor factor $u(xp)$ is understood to be multiplied.  The gluon-line goes into the final-state cut,
and thus has the on-shell
momentum and the physical polarization.  
The quark line with the momentum $k_1$ enters the fragmentation
matrix element and thus the factor for this external line is amputated. 
\label{wti}}
\end{figure}
In order to show the factorization property for the non-pole contribution,
one has to prove that all these gauge-dependent terms
are combined into the matrix elements for the gauge-invariant correlation functions.
To show this, we first note the amplitude with an off-shell
quark external line obeys a WT identity as (see Fig.\,\ref{wti}) 
\begin{eqnarray}
 (k_2-k_1)^\sigma S^{La}_\sigma (k_1,k_2) = T^a S (k_2) + G^a(k_1,k_2), 
 \label{WTI}
\end{eqnarray}
where we have supplied the color index $a$ explicitly.  
The first term $T^aS(k_2)$ results from the sum of the first two terms in the
right-hand-side of Fig. \ref{wti} and the ghost-like term $G^a(k_1,k_2)$
appears due to the off-shellness of the fragmenting
quark.  This ghost-like term, however, can be neglected in our analysis as we will see below.  
The ghost-like term has the following structure, 
\begin{eqnarray}
 G^a(k_1,k_2) &=& {\cal M}^{\mu\lambda} (xp,q,k_1) (k_1-xp-q)_\lambda
  (k_1-xp-q)_\rho
  {\cal P}^{\rho\sigma} (xp+q-k_2) \times \cdots, 
\end{eqnarray}
where ${\cal M}^{\mu\lambda}$ denotes $\gamma^* q\to gq$ scattering amplitude and the polarization tensor
$\cal{P}^{\rho\sigma}$ for the final gluon can be taken as
\begin{eqnarray}
 {\cal P}_{\rho\sigma} (k) = -g_{\rho\sigma} + \frac{k_\rho P_{h\sigma
} + k_\sigma P_{h\rho}}{k\cdot P_h}.  
\end{eqnarray}
Then $G^a(k_1,k_2)$ contains two factors which become zero in the
collinear limit, i.e., 
\begin{eqnarray}
 {\cal M}^{\mu\lambda} (xp,q,\frac{P_h}{z_1})(\frac{P_h}{z_1}-xp-q)_\lambda &=&
  0, \\
 (\frac{P_h}{z_1}-xp-q)_\rho {\cal P}^{\rho\sigma} (xp+q-\frac{P_h}{z_2}) &=& 0.
\end{eqnarray}
Accordingly, the ghost-like term and its first
derivatives with respect to the momenta $k_{1,2\perp}^\alpha$ vanish in the
collinear limit:  
\begin{eqnarray}
 G^a(z_1,z_2) = 0, \quad \del{G^a(k_1,k_2)}{k_{1,2\perp}^\alpha}\cl = 0.
 \label{ghostidentity}
\end{eqnarray} 
%
We now rewrite the WT identity (\ref{WTI}) as 
\begin{eqnarray}
 P_h^\sigma S_\sigma^{La} (k_1,k_2) &=& \frac{1}{1/z_2-1/z_1+i\epsilon}
  \left[ T^a S(k_2) +G^a(k_1,k_2) - (k_{2\perp} - k_{1\perp})^\sigma S_\sigma^{La}
   (z_1,z_2) \right. \nn\\
& & \left. \qquad\qquad -(k_2-k_1)\cdot P_h S^{La}_\sigma (z_1,z_2) w^\sigma  \right],
   \label{WTImod}
\end{eqnarray}
where the sign of the $i\epsilon$-prescription is {\it chosen} so that
this identity reproduces (\ref{GL}) in the collinear limit. 
Noting the relation (\ref{ghostidentity}),
we take the derivative of (\ref{WTImod}) with respect to $k_{1,2\perp}$ and then take the collinear
limit to obtain
\begin{eqnarray}
 \del{S^{La}(k_1,k_2)}{k_{1\perp}^\alpha} \cl
  &=& \frac{1}{1/z_2 - 1/z_1 + i\eps} S_\alpha^{La} (z_1,z_2), 
  \label{WTI1}\\
 \del{S^{La}(k_1,k_2)}{k_{2\perp}^\alpha} \cl
  &=& \frac{1}{1/z_2 - 1/z_1 + i\eps} 
  \left[ T^a \del{S(k_2)}{k_{2\perp}^\alpha} \cl 
   - S_\alpha^{La} (z_1,z_2) \right].\label{WTI2}
\end{eqnarray}
Substituting these relations into the expression of $w^{(b)}$ in (\ref{wb}) and 
integrating over $1/z_1$, one obtains
\begin{eqnarray}
 w^{(b)} &=& i \Oab \int \dZ{1} \dZ{2}
  P\left( \frac{1}{1/z_2 - 1/z_1} \right) \Tr{ {\Delta}_F^{(1)\beta} (z_1,z_2)
  S_\alpha^L (z_1,z_2) } \nn \\
&& - i \Oab \int \dZ{} \Tr{ \left\{ \widehat{\Delta}_F^{(1)\beta} (z) + i\Delta_A^{(1)\beta} (z)
			 + \Delta_{\partial A}^{(1)\beta}(z) \right\}
 \del{S(k)}{k^\alpha}\cl },
 \label{wb3}
\end{eqnarray}
where the single-variable light-cone matrix elements are given by
\begin{eqnarray}
 \widehat{\Delta}_{Fij}^{(1)\beta} (z) &=& \frac{1}{N} \sum_X \int \frac{d\lambda}{2\pi}
  e^{-i\frac{\lambda}{z}} \bra{0} \psi_i (0) \ket{hX} \nn\\
   && \times \bra{hX} \psib_j
  (\lambda w) \left\{ ig \int_{\infty}^{\lambda} d\mu F^{\beta w} (\mu
	       w) \right\} \ket{0}, \\
 \Delta_{Aij}^{(1)\beta} (z) &=& \frac{1}{N} \sum_X \int \frac{d\lambda}{2\pi}
  e^{-i\frac{\lambda}{z}} \bra{0} \psi_i (0) \ket{hX} \bra{hX} \psib_j
  (\lambda w) gA^\beta (\lambda w) \ket{0}, 
\label{DeltaA}\\
 \Delta_{\partial A ij}^{(1)\beta} (z) &=& \frac{1}{N} \sum_X \int
  \frac{d\lambda}{2\pi}
  e^{-i\frac{\lambda}{z}} \bra{0} \psi_i (0) \ket{hX} \nn\\
 && \times \bra{hX} 
  \psib_j (\lambda w) \lpartial^\beta \left\{ ig \int_{\infty}^{\lambda}
				     d\mu A^w (\mu w) \right\} \ket{0}.
				     \label{DeltapartialA}
\end{eqnarray}
Here we observe that
the operators in the matrix elements
appearing in the second term of ({\ref{wb3})
can be regarded as the $O(g)$ part of those in the expansion of $\left[ \psib (\lambda w)
[\lambda w, \infty w] \right] \lpartial^\beta$. 
This term may be 
combined with the matrix element (\ref{D2}) 
which can be regarded as $O(1)$ part of the one for $\left[ \psib (\lambda w)
[\lambda w, \infty w] \right] \lpartial^\beta$. 

By a similar analysis for the diagram (c), it is straightforward to
derive 
\begin{eqnarray}
 w^{(c)} &=& -i \Oab \int \dZ{1} \dZ{2} P \left( \frac{1}{1/z_1 - 1/z_2}
					\right) \Tr{ \Delta_{FR}^{(1)\beta}
 (z_1,z_2) S_\alpha^{R} (z_1,z_2)} \nn \\
 && - i \Oab \int \dZ{} \Tr{ \Delta_{A\partial}^{(1)\beta} (z)
  \del{S(k)}{k^\alpha} \cl}, \label{wc}
\end{eqnarray}
with the light-cone matrix element defined by
\begin{eqnarray}
 \Delta_{A \partial ij}^{(1)\beta} (z) &=& \frac{1}{N} \sum_X \int
  \frac{d\lambda}{2\pi}
  e^{-i\frac{\lambda}{z}} \bra{0} \left\{ ig \int_{0}^{\infty}
				   d\mu A^w (\mu w) \right\} \psi_i (0)
  \ket{hX}
\bra{hX} \psib_j (\lambda w) \lpartial^\beta \ket{0}.
\label{DeltaApartial}
\end{eqnarray}
The operator appearing in this matrix element can be identified as one of $O(g)$ part of 
the gauge-link operator $[\infty w,0]$ in
$\Delta_\partial^\beta (z)$ defined in (\ref{Deltapartial}).   
In this way, one can identify the sum of (\ref{D2}), (\ref{DeltaA}), (\ref{DeltapartialA})
and (\ref{DeltaApartial}) as $\Delta_\partial^\beta (z)$.  
Therefore, taking the sum of (\ref{S}), (\ref{wb3}) and (\ref{wc}), we find that
the sum of $w^{(a)} + w^{(b)} + w^{(c)}$ can be written in a color-gauge-invariant form as 
\begin{eqnarray}
w^{(a)}_{\mu\nu} + w^{(b)}_{\mu\nu} + w^{(c)}_{\mu\nu}
 &=& \int \frac{dz}{z^2}\Tr{\Delta(z)S_{\mu\nu}(z)} + \Oab
 \int \dZ{} {\rm Im} \Tr{
  \Delta_\partial^\beta (z) \del{S_{\mu\nu}(k)}{k^\alpha} \cl } \nn\\
%
&-& \Oab \int \dZ{1} \dZ{2} \PV{ \frac{1}{1/z_2 - 1/z_1} } \nn\\
& & \times
\left\{
{\rm Im} \Tr{ \Delta_F^\beta (z_1,z_2) S_{\mu\nu,\alpha}^L (z_1,z_2)
  } + (\mu \lrarrow \nu) \right\},
\label{wabc}
\end{eqnarray}
where we have dropped the superscripts (0) and (1), and have restored the Lorentz indices $\mu$ and $\nu$
to make the symmetrization for $ S_{\mu\nu,\alpha}^L (z_1,z_2)$ explicit.

Finally, we consider the twist-3 contributions arising
from the diagrams (d,e) in Fig.\,\ref{generic}.
In this case, the corresponding hard part $\St^L_{\mu\nu,\sigma} (k_1,k_2)$
obey a simple Ward identity,
i.e.,
\begin{eqnarray}
 (k_2-k_1)^\sigma \St^L_{\mu\nu,\sigma} (k_1,k_2) = 0, 
\end{eqnarray}
and the resulting factorization
formula is expressed solely in terms of the gauge-invariant $F$-type matrix
element (\ref{Et_F}) as 
\begin{eqnarray}
 w^{(d)}_{\mu\nu} + w^{(e)}_{\mu\nu} &=& - \Oab \int \dZ{1} \dZ{2} 
  \PV{ \frac{1}{1/z_2-1/z_1} } \nn\\
  & &\times
  \left\{
{\rm Im}
 \Tr{\Deltat_F^\beta (z_1,z_2) \St_{\mu\nu,\alpha}^L (z_1,z_2)}
+ (\mu \lrarrow \nu) \right\}.
\label{wde}
\end{eqnarray}
Equations \,(\ref{wabc}) and (\ref{wde}) show that all the contributions
arising from the twist-3 fragmentation functions have definite
factorization property with manifest color gauge-invariance.
We remark that the technique developed in this section can also be used to 
calculate other twist-3 observables which receive non-pole contribution, 
such as the double-spin asymmetry
$A_{LT}$, c.f. \cite{HattaKanazawaYoshida2013}.

\subsection{Electromagnetic gauge-invariance of hadronic tensor}

Here we show that the hadronic tensor obtained in the previous subsection
satisfies the electromagnetic
gauge-invariance.  
We first note that each contribution in (\ref{wabc}) from different twist-3 fragmentation
function does not separately satisfy
electromagnetic current conservation.  However, one
can show the sum of them, $w^{(a)}+w^{(b)}+w^{(c)}$, satisfies this property.
To show this, we note the
twist-3 relation (\ref{EOM2}) allows us to rewrite
the 2-quark correlator (\ref{Deltapartial}) as
\begin{eqnarray}
 \Delta_{ij} (z) &=& \frac{M_N}{2} (\gamma_5\gamma_\alpha\gamma_\lambda)_{ij} 
  \epsilon^{\lambda\alpha wP_h}
  \left[ \int \frac{dz'}{z'^2} \PV{\frac{1}{1/z'-1/z}} {\rm Im}
   \Eh_F(z',z) + {\rm Im} \et(z) \right] \label{2-body2}.
\end{eqnarray}
Inserting this expression and other matrix elements into (\ref{wabc}), we find 
\begin{eqnarray}
 q^\mu \left[ w^{(a)} + w^{(b)} + w^{(c)} \right]_{\mu\nu} &=& -
  \frac{M_N}{2} \epsilon^{\lambda\alpha wP_h}
  \int \frac{dz}{z^2}
   {\rm Im} \et(z) \nn\\
 && \times \Tr{ \gamma_5\gamma_\lambda 
  \left( \gamma_\alpha q^\mu S_{\mu\nu}(z) + \frac{\Slash{P_h}}{z} q^\mu
   \del{S_{\mu\nu}(k)}{k^\alpha}\cl
  \right) } \nn\\
 &-& \frac{M_N}{2} \epsilon^{\lambda\alpha wP_h} \int \frac{dz}{z^2}
   \int \frac{dz'}{z'^2}
  \PV{\frac{1}{1/z'-1/z}} {\rm Im}\Eh_F(z',z) \nn\\
 && \times \Tr{\gamma_5\gamma_\lambda
  \left( \gamma_\alpha q^\mu S_{\mu\nu}(z) + \frac{\Slash{P_h}}{z} q^\mu
   S_{\mu\nu,\alpha}^{L} (z',z) \right)}.
   \label{qmuw}
\end{eqnarray}
In deriving this relation, we have used the fact that
when $q^\mu$ (or $q^\nu$) is contracted with the right virtual-photon-quark vertex of the diagram in Fig. 1 (b),
the result vanishes due to the WT identity in QED.  
From this expression, one finds that (\ref{qmuw}) vanishes if 
the hadronic tensor satisfies
the following relation:
\begin{eqnarray}
 && \epsilon^{\lambda\alpha wP_h}
\left[\gamma_\alpha q^\mu S_{\mu\nu}(z) + \frac{\Slash{P_h}}{z} q^\mu
  \del{S_{\mu\nu}(k)}{k^\alpha}\cl \right] = 0, \label{req1}\\
 && \epsilon^{\lambda\alpha wP_h}
\left[ \gamma_\alpha T^a q^\mu S_{\mu\nu}(z) + \frac{\Slash{P_h}}{z} q^\mu
  S_{\mu\nu,\alpha}^{L a} (z',z)\right] = 0. \label{req2}
\end{eqnarray}
%
%
To prove that (\ref{req1}) is indeed satisfied, we use the WT
identity in QED,
\begin{eqnarray}
 q^\mu \Slash{k} S_{\mu\nu}(k) = 0,
\end{eqnarray}
which holds for the on-shell quark
momentum $k^\mu =(k^+,k^-=\frac{\vec{k}^2_\perp}{2k^+},\vec{k}_\perp)$.  
%
By applying\\
$\epsilon^{\lambda\alpha wP_h}{\partial/\partial k^\alpha}$
to this relation and then taking the collinear
limit $k\to\frac{P_h}{z}$, one obtains the relation (\ref{req1}). 
For the proof of the second relation (\ref{req2}), we rewrite the left-hand-side of (\ref{req2}) as
\begin{eqnarray}
&&\epsilon^{\lambda\alpha wP_h} q^\mu
\left[ \gamma_\alpha T^a  S_{\mu\nu}(z) + \frac{\Slash{P_h}}{z}
  S_{\mu\nu,\alpha}^{L a} (z',z)\right]\nn\\
&&\qquad\qquad\qquad  =\epsilon^{\lambda\alpha wP_h} q^\mu \frac{\Slash{P_h}}{z}
\left[ i \gamma_\alpha T^a { i \over \frac{\Slash{P_h}}{z}+i\epsilon} S_{\mu\nu}(z) + 
  S_{\mu\nu,\alpha}^{L a} (z',z)\right]. 
\end{eqnarray}
One finds that this expression vanishes due to the WT identity in QED if one recalls 
that $ S_{\mu\nu,\alpha}^{L a} (z',z)$ does not contain diagrams in which the collinear
quark and gluon lines merge into a single quark line and it is exactly the first term in $\left[\ \ \right]$, 
and that $\epsilon^{\lambda\alpha wP_h}$
guarantees the physical polarization for the coherent gluon.  
Therefore (\ref{req2}) is also satisfied.
%
We thus find (\ref{wabc}) meets the requirements for the current
conservation.

Regarding the contribution of $\Et_F$ to the hadronic
tensor (\ref{wde}), it is easy to check
\begin{eqnarray}
 q^\mu \left[ w^{(d)} + w^{(e)} \right]_{\mu\nu} = q^\nu \left[
	w^{(d)} + w^{(e)} \right]_{\mu\nu} = 0,
\end{eqnarray}
by using the WT identity in QED.
All in all, the non-pole contribution of the twist-3 fragmentation
function to the hadronic tensor satisfies the electromagnetic
gauge-invariance.

\section{Calculation of $L^{\mu\nu} W_{\mu\nu}$}

\subsection{Single-spin dependent cross section}

With the gauge-invariant factorized expression at hand, we now turn to
derive the Twist-3 fragmentation contribution to the SSA in SIDIS.
To calculate the contraction $L^{\mu\nu}(l, l') W_{\mu\nu}(p,q,P_h)$
in (\ref{diffsigma}),
we introduce the following 4-vectors which are orthogonal to
each other\,\cite{MengOlnessSoper1992,KoikeTanakaYoshida2011}:
\begin{eqnarray}
 T^\mu &=& {1\over Q} \left(q^\mu + 2\xbj p^\mu\right),\nn\\
 X^\mu &=& {1\over q_T} 
  \left\{ {P_h^\mu \over z_f} -q^\mu 
   -\left( 1 + {q_T^2\over Q^2}\right)\xbj p^\mu\right\},\nn\\
 Y^\mu &=& \epsilon^{\mu\nu\rho\sigma}Z_\nu X_\rho T_\sigma ,\nn\\
 Z^\mu &=& -{q^\mu \over Q}. 
\end{eqnarray}
The hadronic tensor $W^{\mu\nu}$ can be expanded in terms of
the six independent tensors ${\cal
V}_k^{\mu\nu}$ ($k=1,\cdots,4, 8, 9$)\,\cite{KoikeTanaka2007M}; 
\begin{eqnarray}
&&{\cal V}_1^{\mu\nu}=X^\mu X^\nu + Y^\mu Y^\nu,\qquad
{\cal V}_2^{\mu\nu}=g^{\mu\nu} + Z^\mu Z^\nu,\nn\\
&&{\cal V}_3^{\mu\nu}=T^\mu X^\nu + X^\mu T^\nu,\qquad
{\cal V}_4^{\mu\nu}=X^\mu X^\nu - Y^\mu Y^\nu,\nn\\
&&{\cal V}_8^{\mu\nu}=T^\mu Y^\nu + Y^\mu T^\nu,\qquad
{\cal V}_9^{\mu\nu}=X^\mu Y^\nu + Y^\mu X^\nu. 
\end{eqnarray}
Using their inverse tensors $\widetilde{\cal V}_k^{\mu\nu}$ given by 
\begin{eqnarray}
&&\widetilde{{\cal V}}_1^{\mu\nu}={1\over 2}(2T^\mu T^\nu
+X^\mu X^\nu + Y^\mu Y^\nu),\qquad
\widetilde{{\cal V}}_2^{\mu\nu}=T^\mu T^\nu,\nn\\
&&\widetilde{{\cal V}}_3^{\mu\nu}=-{1\over 2}(T^\mu X^\nu + X^\mu T^\nu),\qquad
\widetilde{{\cal V}}_4^{\mu\nu}={1\over 2}(X^\mu X^\nu - Y^\mu Y^\nu),\nn\\
&&\widetilde{{\cal V}}_8^{\mu\nu}={-1\over 2}(T^\mu Y^\nu + Y^\mu T^\nu),\qquad
\widetilde{{\cal V}}_9^{\mu\nu}={1\over 2}(X^\mu Y^\nu + Y^\mu X^\nu),
\end{eqnarray}
one obtains  
\begin{eqnarray}
 L_{\mu\nu}W^{\mu\nu} &=& 
  \sum_{k=1,\cdots,4,8,9} \left[ L_{\mu\nu}{\cal V}_k^{\mu\nu} \right]
  \left[W_{\rho\sigma}\widetilde{\cal V}_k^{\rho\sigma}\right] \nn\\
 &=& Q^2\sum_{k=1,\cdots,4,8,9} {\cal A}_k (\phi-\chi)
  \left[W_{\rho\sigma}\widetilde{\cal V}_k^{\rho\sigma}\right], 
  \label{LW}
\end{eqnarray}
where ${\cal A}_k (\phi-\chi) \equiv L_{\mu\nu}{\cal V}_k^{\mu\nu}/Q^2$ is given by
\begin{eqnarray}
 && {\cal A}_1 (\phi) =  1+\cosh^2\psi, \quad {\cal A}_2 (\phi) = -2,
  \quad {\cal A}_3 (\phi) = -\cos\phi\sinh 2\psi, \nn\\
 && {\cal A}_4 (\phi) =  \cos 2\phi\sinh^2\psi, \quad {\cal A}_8 (\phi)
  = -\sin\phi\sinh 2\psi, \quad {\cal A}_9 (\phi) = \sin 2\phi\sinh^2\psi. 
   \label{Ak}
\end{eqnarray}
\begin{figure}[t]
 \begin{center}
 \fig{0.4}{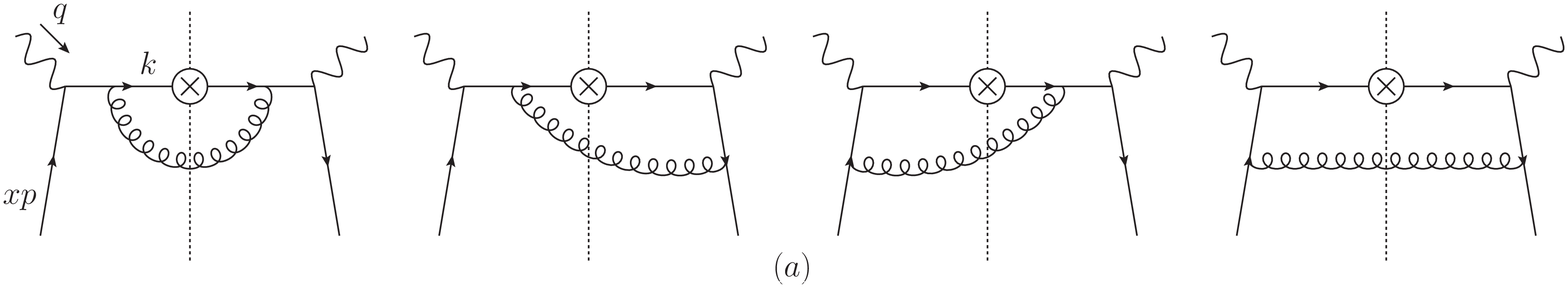}
 \fig{0.4}{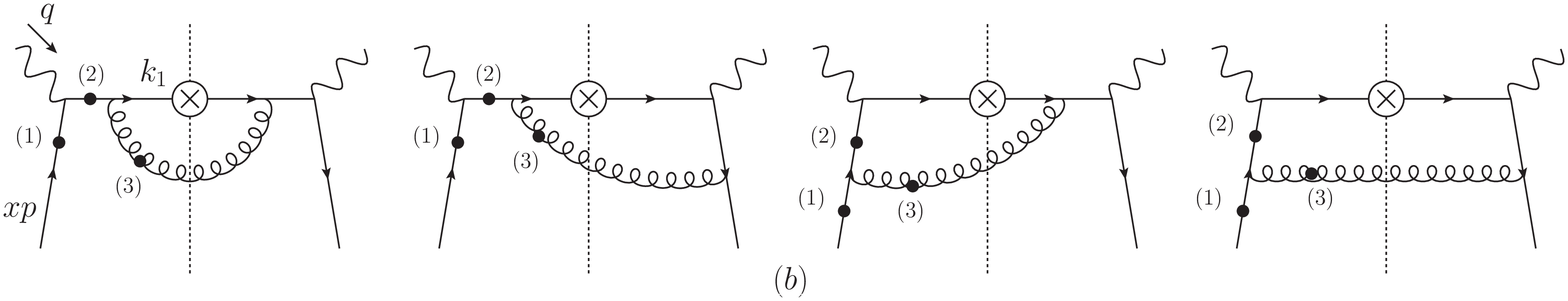}
 \fig{0.4}{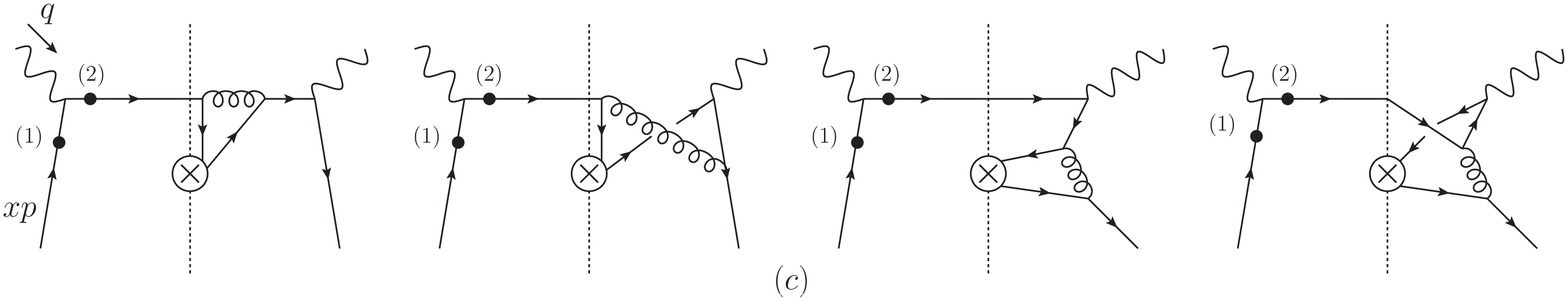}
 \caption{Feynman diagrams giving rise to the contribution of the
  twist-3 fragmentation functions (a) $\eh_{\bar{1}}, \et$, (b) $\widehat{E}_F$ and (c)
  $\widetilde{E}_F$.  For (b) and (c), an extra coherent gluon line coming out of the
  fragmentation insertion $\otimes$ attaches to one of the dots in the
  diagrams. \label{3ff}}
 \end{center}
\end{figure}
By the expansion (\ref{LW}), 
the cross section for the SIDIS process
consists of the five structure functions associated with
${\cal A}_{1,2}$, ${\cal A}_3$, ${\cal A}_4$, ${\cal A}_8$
and ${\cal A}_9$, respectively, which have
different dependencies
on the azimuthal angle $\phi-\chi$.
Calculating the Feynman diagrams shown in Fig.\,\ref{3ff}, the
single-spin-dependent cross section associated with the transversity distribution
and the twist-3 fragmentation functions 
can be obtained to be
\begin{eqnarray}
&& \frac{d^6 \Delta \sigma}{d\xbj dQ^2 dz_f dq_T^2 d\phi d\chi} \nn\\
&=& \frac{\alpha_{em}^2
  \alpha_S M_N}{ 16 \pi^2 \xbj^2 S_{ep}^2 Q^2} \sum_{k=1,\cdots,4,8,9} {\cal
  A}_k {\cal S}_k \int_{x_{min}}^1  \frac{dx}{x} \int_{z_{min}}^{1}
  \frac{dz}{z}
  \delta \left( \frac{q_T^2}{Q^2} - \left(1-\frac{1}{\xh}\right)
	  \left(1-\frac{1}{\zh} \right) \right) \nn\\
&\times&  \sum_a e_a^2 h^a (x)
  \left[ \frac{\eh^a_{\bar{1}}(z)}{z} \Delta\sigmah_k^1 + \frac{d}{d(1/z)}
   \left\{ \frac{{\rm Im} \et^a(z)}{z} \right\}
   \Delta \sigmah_k^2
   + {\rm Im} \et^a(z) \Delta \sigmah_k^3 \right. \nn\\
  && - 2 \left.\int^\infty_z \frac{dz'}{z'^2} 
	  \left\{ \PV{\frac{1}{1/z-1/z'}} {\rm Im}
   \Eh_F^a (z',z) \Delta\sigmah_k^4 \right.\right.\nn\\
 && + \left.\left. z {\rm Im} 
   \Et_F^a (-z',(1/z-1/z')^{-1}) \Delta\sigmah_k^5 \right\} \right], \label{formula}
\end{eqnarray}
where ${\cal A}_k\equiv {\cal A}_k(\phi-\chi)$, ${\cal S}_k \equiv
\sin(\Phi_S-\chi)$ for $k=1,...,4$ and ${\cal S}_k \equiv
\cos(\Phi_S-\chi)$ for $k=8,9$. We have introduced new variables $\xh \equiv \frac{\xbj}{x}$ and
$\zh \equiv \frac{z_f}{z}$.  The symbol $e_a$ denotes the electric charge for quark-flavor $a$ and
the superscript $a$ is supplied to each distribution and fragmentation function.  
The lower limits of the integrations are
given by 
\begin{eqnarray}
 x_{min} &=& \xbj \left( 1 + \frac{z_f}{1-z_f} \frac{q_T^2}{Q^2} \right), \\
 z_{min} &=& z_f \left( 1 + \frac{\xbj}{1-\xbj} \frac{q_T^2}{Q^2} \right). 
\end{eqnarray}
Partonic hard cross sections $\Delta \sigmah_k^i$ ($i=1,\cdots,5$, $k=1,\cdots,4,8,9$)
are the functions of $\xh$, $\zh$ and $\zh'\equiv \frac{z_f}{z'}$ and
are given as follows:
%
%
\begin{eqnarray}
 \Delta\sigmah_1^1 &=& -\frac{4 C_F (-2+5 \zh-3 \zh^2+\xh (4-7 \zh+6 \zh^2))}{q_T
  (1-\zh+\xh (-1+2 \zh))},\\
 \Delta\sigmah_2^1 &=& -\frac{8 C_F \zh}{q_T},\\
 \Delta\sigmah_3^1 &=& -\frac{2 C_F Q ((-1+\zh)^3-2 \xh \zh (4-7 \zh+3
 \zh^2)+\xh^2 (1+7 \zh-14 \zh^2+8 \zh^3))}{q_T^2 \xh \zh (1-\zh+\xh (-1+2 \zh))}
 ,\\
 \Delta\sigmah_4^1 &=&  -\frac{4 C_F Q^2 (-1+\zh) (1-\zh-\xh (-4+\zh) \zh+\xh^2 (-1-3
 \zh+2 \zh^2))}{q_T^3 \xh \zh (1-\zh+\xh (-1+2 \zh))} ,\\
 \Delta\sigmah_8^1 &=& \frac{2 C_F Q (4 \xh (-1+\zh) \zh-(-1+\zh)^2
 (1+\zh)+\xh^2 (1+5 \zh-8 \zh^2+4 \zh^3))}{q_T^2 \xh \zh (1-\zh+\xh (-1+2
 \zh))},\\
 \Delta\sigmah_9^1 &=& -\Delta\sigmah_4^1,
\end{eqnarray}
%
%
\begin{eqnarray}
 \Delta\sigmah_1^2 &=& - \frac{4 C_F (-1+2 \xh) (-1+\zh)}{q_T 
  \left[ 1-\zh+\xh (-1+2\zh) \right]} , \\
 \Delta\sigmah_2^2 &=& 0 , \\
 \Delta\sigmah_3^2 &=& -\frac{4 C_F Q (-1+2 \xh) (-1+\zh)}{q_T^2 
  \left[ 1-\zh+\xh (-1+2 \zh) \right]} , \\
 \Delta\sigmah_4^2 &=& - \frac{4 C_F Q^2 (-1+2 \xh) (-1+\zh)}{q_T^3 
  \left[ 1-\zh+\xh (-1+2 \zh) \right]} , \\
 \Delta\sigmah_8^2 &=& - \Delta\sigmah_3^2 , \\
 \Delta\sigmah_9^2 &=& - \Delta\sigmah_4^2 , 
\end{eqnarray}
%
%
\begin{eqnarray}
 \Delta\sigmah_1^3 &=& \frac{4 C_F (-1+3 \zh)}{q_T} , \\
 \Delta\sigmah_2^3 &=& \frac{8 C_F \zh}{q_T} , \\
 \Delta\sigmah_3^3 &=& 
  \frac{2 C_F Q \left[-(-1+\zh)^2+\xh (1-3 \zh+4 \zh^2)\right]}{q_T^2 \xh \zh} , \\
 \Delta\sigmah_4^3 &=& \frac{4 C_F Q^2 
  \left[-1+\zh+\xh (1-\zh+\zh^2)\right]}{q_T^3 \xh \zh} , \\
 \Delta\sigmah_8^3 &=& \frac{2 C_F Q \left[ 1-\zh^2-\xh(1-\zh+2\zh^2)
			       \right]}{q_T^2 \xh\zh} , \\
 \Delta\sigmah_9^3 &=& -\Delta\sigmah_4^3 ,
\end{eqnarray}
%
%
\begin{eqnarray}
 \Delta\sigmah^4_1 &=& \frac{1}{N q_T} 
 \left[
  \frac{-1+\xh+4 \zh-3 \xh \zh+3 (-1+2 \xh) \zh^2}{-1+\xh+\zh} 
  - \frac{2 (-1+\xh+\zh-\xh \zh)}{(-1+\xh+\zh-2 \xh \zh) (\zh-\zh')} \right. \nn\\
 && \left. + \frac{\xh (-1-6 (-1+\zh) \zh)}{\zh'}
  - \frac{6 (-1+\xh) \xh (-1+\zh) \zh^2}{(-1+\xh+\zh) (\xh \zh+\zh'-(\xh+\zh) \zh')}
 \right] \nn\\
 &+& \frac{N}{q_T} 
  \left[ \frac{1-4 \zh+3 \zh^2+\xh (-1+3 \zh-6 \zh^2)}{-1+\xh+\zh}
   - \frac{2 (-1+2 \xh) (-1+\zh) \zh}{(-1+\xh+\zh-2 \xh \zh) (\zh-\zh')} \right. \nn\\
 && \left. - \frac{6 \xh (-1+\zh) \zh^3}{(-1+\xh+\zh) (\xh \zh+\zh'-(\xh+\zh) \zh')} 
  \right] , \\
 \Delta\sigmah^4_2 &=& \frac{2 \zh (2 \xh^2 (-1+\zh) (\zh-\zh')-(-1+\zh) \zh'^2+\xh \zh' (-2-\zh'+\zh (3-2 \zh+2
 \zh')))}{N q_T \zh' (-\xh \zh+(-1+\xh+\zh) \zh')} \nn\\ 
 &+& \frac{2 N \zh (-\xh (-1+2 \zh) (\zh-\zh')+\zh'-\zh \zh')}{ q_T (\xh
  \zh - (-1+\xh+\zh) \zh')} , \\
 \Delta\sigmah^4_3 &=& \frac{Q}{2 N q_T^2} 
  \left[ \frac{(-1 + \zh)^3 - 2 \xh (-1 + \zh) (1 - 2 \zh + 4 \zh^2) + 
    \xh^2 (-1 + \zh (3 + 8 (-1 + \zh) \zh))}{\xh \zh (-1 + \xh + \zh)}
  \right. \nn\\
 && - \frac{4 (-1 + \xh + \zh - \xh \zh)}{(-1 + \xh + \zh - 2 \xh
     \zh) (\zh - \zh')} 
    - \frac{2 (-1 + 2 \xh) (-1 + \zh) (-1 + 2 \zh)}{\zh'} \nn\\
 && \left. - \frac{ 4 (-1 + \xh) (-1 + \zh) \zh (1 - \zh + \xh (-1 + 2
    \zh))}{(-1 + \xh + \zh) (\xh \zh + \zh' - (\xh + \zh) \zh')} \right]
 \nn\\
 &+& \frac{NQ}{2 q_T^2 \zh} 
  \left[ 1 - \frac{(-1 + \zh)^2}{\xh} + 
   \zh \left\{ -3 + 8 (-1 + \zh) \zh \left( -1 + \frac{1}{(-1 + 2 \zh)
				     (\zh - \zh')}
      \right) \right\} \right. \nn\\
 && \left. + \frac{4 (-1 + \zh) \zh^2}{(-1 + 2 \zh) (-1 + \xh + \zh - 2 \xh
     \zh) (\zh -  \zh')} 
      + \frac{4 (-1 + \zh) \zh^3 (-1 + 2 \zh')}{-\xh \zh + (-1 + \xh + \zh)
      \zh'} \right], \\
 \Delta\sigmah^4_4 &=& \frac{1}{N q_T} 
  \left[ 
   \frac{-(-1+\zh)^2+\xh (1+\zh (-1+2 \zh))}{-1+\xh+\zh}
  - \frac{(1+2 (-1+\xh) \xh) (-1+\zh) \zh}{(-1+\xh) \zh'} \right. \nn\\
 && \left. - \frac{2 \xh \zh}{(-1+\xh+\zh-2 \xh \zh) (-\zh+\zh')}
  + \frac{\zh (-(-1+\zh)^2+2 \xh (-1+\zh)^2+\xh^2 (-1-2 (-1+\zh)
  \zh))}{(-1+\xh+\zh) (\xh \zh+\zh'-(\xh+\zh) \zh')} \right] \nn\\
 &+& \frac{N}{q_T} 
  \left[ \frac{(-1+\zh)^2+\xh (-1+\zh-2 \zh^2)}{-1+\xh+\zh} 
   - \frac{2 \xh (-1+2 \xh) \zh^2}{(-1+\xh) (-1+\xh+\zh-2 \xh \zh) (\zh-\zh')}
  \right. \nn\\
 && \left. + \frac{\zh^2 (-(-1+\zh)^2+2 \xh (-1+\zh)^2+\xh^2 (-1-2
     (-1+\zh) \zh))}{(-1+\xh) (-1+\xh+\zh) (\xh \zh+\zh'-(\xh+\zh)
     \zh')} \right] , \\
 \Delta\sigmah^4_8 &=& \frac{Q}{2 N q_T^2} 
  \left[
   - \frac{(1+\zh) (1-\zh+\xh (-1+2 \zh))}{\xh \zh} 
   - \frac{4 (-1+\xh) (-1+\zh)}{(-1+\xh+\zh-2 \xh \zh) (\zh-\zh')} \right. \nn\\
 && \left. + \frac{2-6 \zh+4 \zh^2}{\zh'} 
     + \frac{4 (-1+\xh) (-1+\zh) \zh}{\xh \zh+\zh'-(\xh+\zh) \zh'}
  \right]\nn\\
 &+& \frac{N Q}{2 q_T^2 \zh} 
  \left[ -1+\zh+2 \zh^2+ \frac{1-\zh^2}{\xh} 
   + \frac{4 (-1+2 \xh) (-1+\zh) \zh^2}{(-1+\xh+\zh-2 \xh \zh) (\zh-\zh')}
+ \frac{4 (-1+\zh) \zh^3}{\xh \zh+\zh'-(\xh+\zh) \zh'}
    \right] , \\ 
 \Delta\sigmah^4_9 &=& \frac{1}{N q_T} 
  \left[
   -1 + \zh \left\{ -1 - \frac{2 \xh}{(-1+\xh+\zh-2 \xh \zh) (\zh-\zh')} \right. \right. \nn\\
 && \left. \left. + \frac{(-1+2 \xh) (-1+\zh)}{(-1+\xh) \zh'} 
	    + \frac{-1+\xh+\zh-2 \xh \zh}{-\xh \zh+(-1+\xh+\zh) \zh'} \right\} \right]
  \nn\\
 &+& \frac{N}{q_T} 
  \left[ 
   1 \right. \nn\\ 
 && \left. +\zh \left\{ 1 + \frac{\zh}{(-1+\xh) (1-\zh+\xh (-1+2 \zh))} \left( \frac{2 (1-2 \xh) \xh}{\zh-\zh'}
   + \frac{(-1+\xh+\zh-2 \xh \zh)^2}{\xh \zh+\zh'-(\xh+\zh) \zh'} \right)
	\right\} \right] ,
\end{eqnarray}
%
%
\begin{eqnarray}
 \Delta\sigmah_1^5 &=& \frac{\xh (-1 + \zh) [1 + 6 (-1 + \zh) \zh] }{N
 q_T \zh } F , \\
 \Delta\sigmah_2^5 &=& \frac{4 \xh (-1+\zh)^2 }{N q_T} F, \\
 \Delta\sigmah_3^5 &=& \frac{Q (-1 + 2 \xh) (-1 + \zh)^2 (-1 + 2 \zh)
 }{N q_T^2 \zh } F , \\
 \Delta\sigmah_4^5 &=& \frac{Q^2 (1 + 2 (-1 + \xh) \xh) (-1 + \zh)^3 }{N
 q_T^3 \xh \zh } F , \\
 \Delta\sigmah_8^5 &=& - \frac{Q (-1 + \zh)^2 (-1 + 2 \zh) }{
 N q_T^2 \zh } F , \\
 \Delta\sigmah_9^5 &=& - \frac{Q^2 (-1 + 2 \xh) (-1 + \zh)^3 }{N q_T^3
 \xh \zh } F, 
\end{eqnarray}
where $F$ is a dimensionless factor given by 
\begin{eqnarray}
 F &=& \frac{ \xh \zh^3 + 
     \zh (\xh - (-1 + \zh)^2 - 3 \xh \zh) \zh' + (-1 + \xh + 
        \zh)^2 \zh'^2 }{ \zh' [ (-1+\zh)
 \zh - (-1+\xh+\zh) \zh'] [-\xh \zh+(-1+\xh+\zh) \zh'] }. 
\end{eqnarray}

\subsection{Azimuthal asymmetries and their asymptotic behavior at small-$q_T$}

Using the obtained cross-section formula (\ref{formula}), we investigate
the small-$q_T$ behavior of the azimuthal asymmetries, which becomes important 
to check the consistency with the description based on the TMD
factorization approach. 

We now define new angle variables as, 
\begin{eqnarray}
 \phi-\chi=\phi_h, \quad \Phi_S-\chi=\phi_h-\phi_S, 
\end{eqnarray}
where $\phi_h$ and $\phi_S$ are, respectively, the azimuthal angles of the hadron
plane and the nucleon's spin vector measured from the lepton
plane.  With these variables, the polarized cross-section is now recast into
\begin{eqnarray}
 \frac{d^6 \Delta \sigma}{d \xbj dQ^2 dz_f dq_T^2 d\phi d\chi} &=&
  \sin(\phi_h-\phi_S) \left[ {\cal F}_1 + {\cal F}_2\cos\phi_h + {\cal
		       F}_3\cos 2\phi_h \right] \nn\\
 && + \cos(\phi_h-\phi_S) \left[ {\cal F}_4\sin\phi_h+{\cal
			   F}_5\sin2\phi_h \right] \nn\\
 &=& F^{\sin(\phi_h-\phi_S)} \sin(\phi_h-\phi_S)  
 + F^{\sin(2\phi_h-\phi_S)} \sin(2\phi_h-\phi_S) 
 + F^{\sin\phi_S} \sin\phi_S \nn\\
 && + F^{\sin(3\phi_h-\phi_S)} \sin(3\phi_h-\phi_S) 
   + F^{\sin(\phi_h+\phi_S)} \sin(\phi_h+\phi_S) \label{azimuthal}, 
\end{eqnarray}
where the structure functions with different azimuthal dependencies are
given by
\begin{eqnarray}
 && F^{\sin(\phi_h-\phi_S)} = {\cal F}_1, \quad F^{\sin(2\phi_h-\phi_S)}
  = \frac{{\cal F}_2+{\cal F}_4}{2} , \quad F^{\sin\phi_S}
  = \frac{-{\cal F}_2+{\cal F}_4}{2} , \nn\\
 && F^{\sin(3\phi_h-\phi_S)} = \frac{{\cal F}_3+{\cal F}_5}{2}
			       , \quad F^{\sin(\phi_h+\phi_S)} =
 \frac{-{\cal F}_3+{\cal F}_5}{2} .
\end{eqnarray}
In order to see the asymptotic behavior of the structure functions 
at small-$q_T$, we need the expression for the $\delta$-function 
in the $q_T\to 0$ limit; 
\begin{eqnarray}
 \delta \left( \frac{q_T^2}{Q^2} - \left(1-\frac{1}{\xh}\right)
	  \left(1-\frac{1}{\zh} \right) \right) {\rightarrow}
 \xh\zh \left[ \frac{\delta(\xh-1)}{(1-\zh)_+} +
	 \frac{\delta(\zh-1)}{(1-\xh)_+} + \delta(\xh-1)\delta(\zh-1)
	 \ln \frac{Q^2}{q_T^2} \right].
\end{eqnarray}
Using this form in (\ref{formula}), we find 
$F^{\sin(\phi_h-\phi_S)} \sim 1/q_T$,
$F^{\sin(2\phi_h-\phi_S)} \sim 0$,
and $F^{\sin(3\phi_h-\phi_S)}\sim 0$ for the contributions from the twist-3 fragmentation functions,
and thus they are suppressed at small-$q_T$ compared
with the contribution from the quark-gluon correlation functions inside the
transversely polarized nucleon
\cite{EguchiKoikeTanaka2007,KoikeTanaka2009,BacchettaBoerDiehlMulders2008}.   
On the other hand twist-3 fragmentation function gives a
leading contribution to the other two structure functions as
\begin{eqnarray}
 && F^{\sin\phi_S} \simeq \frac{F_0}{Qq_T^2} \sin 2\psi \sum_a e_a^2
  \int\frac{dx}{x}  h^a(x) \int\frac{dz}{z}
  A_1 \delta(\xh-1) , \label{sinPS}\\
 && F^{\sin(\phi_h+\phi_S)} \simeq - \frac{F_0}{q_T^3}
  \sinh^2\psi \sum_a e_a^2 \int\frac{dx}{x} h^a(x) \int\frac{dz}{z}
 \left[ A_2 \delta(\xh-1) + B_2 \delta(\zh-1) \right] ,\label{Col}
\end{eqnarray}
with $F_0=\frac{\alpha_{em}^2 \alpha_S M_N}{16\pi^2 \xbj^2 S_{ep}^2}$.  
Here $A_1$, $A_2$ and $B_2$ are, respectively, given by
\begin{eqnarray}
 A_1 &=& 2C_F \left[ \frac{\eh_{\bar{1}}^a(z)}{z} 
	       \left( 3\zh-\frac{2}{(1-\zh)_+} \right) 
+ 2\frac{d}{d(1/z)} \left\{ \frac{{\rm Im}\et^a(z)}{z} \right\} + {\rm
			     Im}\et^a(z) \frac{\zh(-1+3\zh)}{(1-\zh)_+}
			    \right] \nn\\
 &-& 2 \int_z^\infty \frac{dz'}{z'^2} \left[ \PV{ \frac{1}{1/z-1/z'} } 
  {\rm Im} \Eh_F^a (z',z) \right. \nn\\
 && \times \frac{\zh}{2} 
  \left[ \frac{1}{N}
   \left( \frac{2(2\zh-1)}{\zh'}+\frac{1+\zh}{(1-\zh)_+} \right) 
  + N \left(
       \frac{4\zh}{1-\zh'}-\frac{4}{\zh-\zh'}-\frac{1+\zh}{(1-\zh)_+}
      \right) \right] \nn\\
 && + \left. z{\rm Im}\Et_F^a (-z',(1/z-1/z')^{-1})
       \frac{(1-\zh)(2\zh-1)(\zh-\zh')}{N(1-\zh+\zh')\zh'} \right], \\
 A_2 &=& 4C_F \left[ \frac{\eh_{\bar{1}}^a(z)}{z} \zh + \frac{d}{d(1/z)}
	   \left\{ \frac{{\rm Im} \et^a(z)}{z} \right\} + {\rm Im} \et^a(z)
	   \frac{\zh^2}{(1-\zh)_+} \right] \nn\\
 &-&2 \int_z^\infty \frac{dz'}{z'^2} \left[ \PV{ \frac{1}{1/z-1/z'} } 
  {\rm Im} \Eh_F^a (z',z) \zh
  \left[
   N \left(\frac{\zh}{1-\zh'}-\frac{2}{\zh-\zh'}\right)  -
   \frac{1-\zh}{N\zh'} \right]
  \right. \nn\\
 && \left. + z {\rm Im} \Et^a_F(-z',(1/z-1/z')^{-1})
  \frac{(1-\zh)^2 (\zh-\zh')}{N\zh'(-1+\zh-\zh')} \right], \\
 B_2 &=& 4C_F {\rm Im}\et^a(z_f) \left[ \frac{\xh}{(1-\xh)_+} + \delta(\xh-1)
      \ln\frac{Q^2}{q_T^2} \right].
\end{eqnarray}
Note the terms proportional to $\delta(\zh-1)$ in (\ref{sinPS})
cancel due to the twist-3 relation (\ref{EOM2}).  
Twist-3 quark-gluon correlation function in the nucleon
gives rise to these structure functions as
$F^{\sin(\phi_S)}_{\rm distribution}\sim 1/q_T^2$
and $F^{\sin(\phi_h+\phi_S)}_{\rm distribution}\sim 1/q_T$, which may be compared with the above result
(\ref{sinPS}) and (\ref{Col}).  Thus, for the $\sin(\phi_S)$-asymmetry, 
two contributions are equally important.  
The obtained asymptotic part (\ref{Col}) for the Collins azimuthal
asymmetry does not agree with that derived in the previous study \cite{YuanZhou2009} in which the
authors did not include
the contribution from $\eh_{\bar{1}}$ and $\Et_F$.  
We emphasize that the inclusion of $\eh_{\bar{1}}$-contribution is crucial to
guarantee the electromagnetic gauge-invariance, as we saw in Sec. 3.4. 

For a confirmation of the matching/mismatching between the collinear twist-3
approach and the TMD factorization approach for all structure functions, we
need to know the asymptotic behavior of the relevant TMD functions in its high
transverse-momentum region.   We will investigate this issue in the future publication.  

\section{Summary}

In this paper, we have calculated the contribution of the twist-3
fragmentation function to the SSA in SIDIS.  We have established the
collinear twist-3 formalism in the Feynman gauge to derive the single-spin
dependent cross section formula.  
There, the relations among hard parts based on the
WT identities play a crucial role in reorganizing the matrix
elements into the color gauge-invariant ones.
We have also shown the obtained twist-3 hadronic tensor satisfies the
electromagnetic gauge-invariance.
Together with the formalism for the pole contributions
\cite{EguchiKoikeTanaka2007,BeppuKoikeTanakaYoshida2010}, 
present work completes the theoretical formalism to calculate the twist-3 cross sections for 
$p_T$-dependent processes
within the framework
of the collinear factorization in QCD. 
Using the complete cross-section formula, we have derived the asymptotic
behavior of the azimuthal asymmetries in the small-$q_T$ region.

\section*{Acknowledgments}

We thank Andreas Metz, Kazuhiro Tanaka, Feng Yuan and Jian Zhou for useful discussions.
The work of K.K. is supported by the Grand-in-Aid for
Scientific Research (No.24.6959) from the Japan Society
of Promotion of Science.
The work of Y.K. is supported in part by the Grant-in-Aid for
Scientific Research
(No.23540292) from the Japan Society of Promotion of Science.  


\end{document}